# Machine Learning and Meta-Analysis Approach to Identify Patient Comorbidities and Symptoms that Increased Risk of Mortality in COVID-19

**Sakifa Aktar [1, #], Ashis Talukder [2, #], Md. Martuza Ahamad [1, #], A. H. M. Kamal [3], Jahidur Rahman Khan [4], Md. Protikuzzaman [1], Nasif Hossain [5], Julian M.W. Quinn [6,7], Mathew A. Summers [6,8], Teng Liaw[9], Valsamma Eapen[10, *], Mohammad Ali Moni [9,10, *]**

1. Department of Computer Science and Engineering, Bangabandhu Sheikh Mujibur Rahman Science and Technology University, Gopalganj-8100, Bangladesh.
2. Statistics Discipline, Khulna University, Khulna-9208, Bangladesh.
3. Dept. of Computer Science and Engineering, Jatiya Kabi Kazi Nazrul Islam University, Bangladesh.
4. Health Research Institute, University of Canberra, ACT, Australia.
5. School of Tropical Medicine and Global Health, Nagasaki University, Japan.
6. The Garvan Institute of Medical Research, Healthy Ageing Theme, Darlinghurst, NSW, Australia.
7. Royal North Shore Hospital SERT Institute, Division of Surgery and Anesthesia, St Leonards, NSW, Australia.
8. St Vincent's Clinical School, University of New South Wales, Faculty of Medicine, Sydney, Australia.
9. World Health Organization (WHO) Centre on eHealth, School of Public Health and Community Medicine, Faculty of Medicine, University of New South Wales, Sydney, Australia.
10. School of Psychiatry, Faculty of Medicine, University of New South Wales, Sydney, Australia.

[#] Authors contributed equally to this work

* Corresponding author
Dr Mohammad Ali Moni (m.moni@unsw.edu.au)
World Health Organization (WHO) Centre on eHealth, School of Public Health and Community Medicine, Faculty of Medicine, University of New South Wales, Sydney, Australia
&
School of Psychiatry, Faculty of Medicine, University of New South Wales, Sydney, Australia

Prof Dr Valsamma Eapen (v.eapen@unsw.edu.au)
School of Psychiatry, Faculty of Medicine, University of New South Wales, Sydney, Australia

**Abstract**

Caring for people suffering from COVID-19 is a significant global challenge. Many individuals infected have pre-existing conditions that interact to increase symptom severity and mortality risk. To assess the interaction of patient comorbidities with COVID-19 we performed a meta-analysis of the published global literature, and machine learning predictive analyses using an aggregated COVID-19 global dataset. Meta-analysis identified chronic obstructive pulmonary disease (COPD), cerebrovascular disease (CEVD), cardiovascular disease (CVD), type 2 diabetes, malignancy, and hypertension as most significantly associated with COVID-19 severity. Machine learning classification using novel aggregated cohort data found COPD, CVD, CKD, type 2 diabetes, malignancy, hypertension, and asthma as the most significant features for classifying deceased versus survived in COVID-19. Symptom-comorbidity interaction analysis found the combinations of Pneumonia-Hypertension, Pneumonia-Diabetes, and Acute Respiratory Distress Syndrome with Hypertension had the most significant effects on mortality. These results highlight patients most at risk of COVID-19 related severe morbidity and mortality.

## Introduction

As of the end of May 2020, over 6 million cases of SARS-CoV-2 infection have been confirmed globally, and over 3,696,000 deaths attributed to the associated disease, COVID-19.[1] Asymptomatic human to human spread remains a challenging aspect of the viral containment effort, unlike previous pandemic coronaviruses SARS and MERS, which showed co-occurrence of symptoms with infectiousness.[2] COVID-19 epidemiological data suggests elderly people are most at risk of developing severe symptoms,[3] although severe symptoms and mortality occur in all age groups. The more prominent symptoms include high fever, cough and sputum production, headache, hemoptysis and diarrhea, and as the infection worsens, an acute respiratory distress syndrome can develop that requires intensive care management. Identifying those most at risk of severe symptoms and death remains a research priority to aid early and appropriate allocation of resources and targeted patient management. As more population data is released, predictive analytic methods may be able to provide such information for patients based on their clinical characteristics.

Reports are emerging that many of the patients most affected by COVID-19 also present with significant comorbidities. A recent study by Richardson *et al.*[4] describing 5,700 confirmed COVID-19 cases reported that many of these patients were suffering from hypertension (56.6%), obesity (41.7%) or type 2 diabetes (33.8%) at the time of their infection; greater than their respective prevalence in the population, which suggests a link to SARS-CoV-2 effects on metabolic and vascular systems. This indicates that the comorbidities an individual has may provide crucial prognostic information if SARS-CoV-2 infection co-occurs. There is also data emerging that suggests significant heterogeneity in disease presentation. Xu *et al.*[5] described clinical characteristics (including laboratory and chest radiography data) from 62 Chinese COVID-19 patients that differed from those described by Guan *et al.* in another Chinese region.[6] The reasons for this heterogeneity in presentations remain unclear, but the relative incidence of comorbidities (and other clinical features) in different patient cohorts provide one explanation. The nature and strength of comorbidity interaction with COVID-19 may also provide important clues to the mechanisms of their interaction and how this may be countered.

To address these issues, we used three approaches to analyze the currently available clinical information. Firstly, we conducted a meta-analysis of available retrospective cohort studies of COVID-19 patient data that focused on comorbidity and selected clinical features. Secondly, we also obtained and aggregated a novel COVID-19 dataset from 4,81,289 patients from across 141 different countries[7], and identified significant comorbidity associations. Thirdly, we applied machine learning algorithms to this novel aggregated data to classify comorbidities with mortality. These three approaches enabled us to thoroughly assess the comorbidities and clinical features most significantly associated with mortality in COVID-19 patients.

## Results

### Meta-analysis of published clinical reports of COVID-19 disease

Initially, our meta-analysis search terms identified a total of 195 relevant articles. From these articles, we excluded 96 duplicate references and considered the remaining 99. By careful screening of the title and abstract, we excluded 34 articles based on the criteria noted above (e.g., we did not include case reports, review reports) and we only considered full-text papers that examined comorbidity and clinical symptoms on COVID-19 patients; these are listed in Table 1. Finally, for the remaining articles, we reviewed the full text and removed a further 36 studies because they were either reviews or editorials lacking clinical details. Twenty-six articles eventually met the inclusion criteria for our meta-analysis. A flow-diagram of literature screening is shown in Figure 1.

A total of 13,400 COVID-19 patients from twenty-six studies [4-6, 18-40] were thus included in our meta-analysis. Most of the studies were conducted in China (24), one was from the USA, and another was from Italy. The mean age of the full sample was 54.5 years, with 8,149 (60.81%) males and 39.19% females (Table 1). Of these, there were 2,964 patients (22.11%) who developed a severe condition or were admitted to the ICU or had died (Table 1). Note that, for calculating the prevalence we considered the full data set from all the 26 publications. However, due to lack of information (patients were not stratified based on the degree of severity), we considered only 11 publications in the analysis to assess the effect of symptoms and comorbidities on COVID-19 disease severity or death.

The results of our meta-analysis show the dominant symptomology in COVID-19 disease. Fever (typically defined by a body temperature above 38.5°C though sometimes not precisely defined) was the most prevalent feature (88.26%, 95% CI 81.31, 92.84%) (Table 2). The next most common significant symptom was persistent cough (63.68%, 95% CI 57.49, 69.45), followed by excessive fatigue (40.48%, 95% CI 34.49, 48.77), dyspnea (26.49%, 95% CI 18.50, 36.39), anorexia (21.92%, 95% CI 13.50, 33.56), myalgia (21.01%, 95% CI 15.50, 27.82), headache (9.84%, 95% CI 7.38, 13.00), diarrhea (7.60%, 95% CI 4.89, 11.63) and nausea (6.50%, 95% CI 3.10, 13.10) (as shown in Table 2).

Hypertension (23.41%, 95% CI 17.63, 30.63) was the most prevalent comorbidity observed among COVID-19 patients, followed by diabetes (11.84%, 95% CI 8.27, 18.14), CVD (10.00%, 95% CI 7.68, 12.93), malignancy (4.09%, 95% CI 3.18, 5.24), cerebrovascular disease (CEVD; 3.23%, 95% CI 2.02, 5.13), chronic obstructive pulmonary disease (COPD 3.18%, 95% CI: 2.33, 4.34), chronic kidney disease (CKD; 2.78%, 95% CI 1.74, 4.41) and chronic liver disease (CLD 2.50%, 95% CI 1.51, 4.11) (Table 3); prevalence of smoking was 8.83% (95% CI 4.19, 17.69) (Table 3). Note that prevalence was estimated using a random effects model, and significant ($p < 0.05$) high heterogeneities were observed for the estimates, with $I^2$ ranging from 79 to 99% (see Table 3).

Table 4 shows the meta-analysis results of the association between symptoms as well as comorbidities in severe and non-severe patients from those articles where severity, ICU support requirement or death were reported. When clinical symptoms were stratified according to patient severity, higher odds of dyspnea (OR= 2.43, 95% CI 1.52, 3.89) were observed in the severe symptom group. Thus, COVID-19 patients with dyspnea have more than two fold increase in risk of developing severe symptoms. The odds of fever (OR= 1.04, 95% CI: 0.85, 1.28), cough (OR 1.12, 95% CI 0.91, 1.38), fatigue (OR 1.14, 95% CI 0.96, 1.36), anorexia (OR 1.56, 95% CI 0.93, 2.62), myalgia (OR 0.78, 95% CI 0.54, 1.13), headache (OR 1.04, 95% CI 0.69, 1.56), diarrhea (OR 1.14, 95% CI 0.81, 1.61) and nausea (OR 0.93, 95% CI 0.58, 1.47) were also found to be higher in COVID-19 patients with severe symptoms.

COPD was found to be the comorbidity feature most significantly associated with high disease severity since the odds ratio of COPD (OR 4.76, 95% CI 2.69, 8.39) was the highest among all other comorbidities and conditions that were considered. The next most significant comorbidity (or condition) relating to disease severity was CEVD (OR 4.54, 95% CI 2.29, 8.99) followed by CVD (OR 3.46, 95% CI 2.05, 5.87), CKD (OR 3.22, 95% CI 1.70, 6.10), type II diabetes (OR 2.08, 95% CI 1.39, 3.10), malignancy (OR 2.04, 95% CI 1.02, 4.07), hypertension (OR 1.81, 95% CI 1.49, 2.20) and smoking (OR 1.74, 95% CI 1.25, 2.42).

## Publication bias

In parallel to the meta-analysis of data, we also conducted an analysis of publication bias for all symptoms and comorbidities. Table 4 shows the results of possible publication biases, which were assessed using funnel plots and Egger's testing (for details, see, Supplementary Figure 3). The results of the Egger's test ($p > 0.05$) suggest that except for the symptom of anorexia, there were no significant publication biases seen in the variables analyzed.

## Clinical characteristics of patients in aggregated recently generated COVID-19 patient datasets

Following our meta-analysis of the published literature, we also sought to assess recent COVID-19 clinical case data available from open-source online repositories; this allowed us to apply additional novel predictive machine learning methods to COVID-19 data complementing our meta-analysis of the published literature. Data were obtained from two different large data repositories and processed as detailed in the methods section. Following filtering for case data to include only cases with sufficiently detailed clinical information, as well as case mortality information, we obtained a total of 1,143 patient cases for analysis. Table 5 displays summary statistics of these 1,143 patients stratified by survival/mortality outcomes. The analysis found that of the 1,143 patients, 86.61% had no comorbidities, whereas 5.34% and 7.87% of patients had only one or more than one comorbidity, respectively. The most common coexisting comorbidities were hypertension (8.66%), diabetes (7.44%), cardiovascular disease (3.5%), and kidney disease (1.75%). In contrast, malignancy of any kind (0.87%), asthma (0.87%), COPD (0.61%), chronic lung disease (0.61%), cerebrovascular disease (0.44%), surgical history (0.26%), neurodegenerative disease (0.17%), infectious disease (0.17%), and liver disease (0.17%) were found to be far less likely to co-occur with COVID-19 in this dataset. Analyzing this data for clinical symptomatology found that the most common clinical presentation of patients with COVID-19 was fever (14.17%) followed by cough (12.42%), pneumonia (6.47%), acute respiratory distress symptoms (5.69%), dyspnea (3.06%), fatigue (2.19%), septic shock (1.49%), headache (0.96%), myalgia (0.79%), diarrhea (0.61%) and nausea (0.26%).

Table 5 also shows the status of patients who were deceased. The selected 1143 patients included 319 (27.91%) deceased, of which 32.60% were female and 61.76% were male. The median age of the deceased patients was 51 years and IQR of 36 to 66 years. A majority of patients (67.08%) had no comorbidities in this dataset. Only 10.97% of patients had one comorbidity, while 21.94% had more than one comorbidity. In the deceased patient subgroup, the rate of comorbidities was significantly higher than survived patients. The comorbidities most frequently seen in COVID-19 patients that did not survive their infection included type 2 diabetes (19.12%), cardiovascular disease (6.27%), and kidney disease (4.08%). However, while the other comorbidities we studied (see Table 5) were less frequently observed in COVID-19 patients, when they did co-occur, they did so only in patients who had died (Table 5). Descriptive analysis of the symptoms in the deceased COVID-19 patients found that the most significant symptoms seen in the deceased patients were pneumonia (21.32%), fever (12.85%), cough (11.60%), acute respiratory distress symptom (9.72%) and septic shock (4.70%) (Table 5).

## Supervised machine learning identifies the most significant COVID-19 comorbidities

To predict significant COVID-19 comorbidities, and to compare with our meta-analysis of the published literature, we designed and performed a machine learning analysis of our 1,143 patient's datasets. We applied six different machine learning algorithmic approaches (Random Forest, Decision Tree, GBM, XGB, SVM and LGBM) to identify the best predictors of COVID-19 patient mortality among the comorbidities and symptoms. We achieved a regression accuracy of > 80% in all six approaches to comorbidity and mortality; specifically, that was 83% for Decision Tree, 84% for GBM, and 86% for XGB, 87% for Random Forest and SVM, and 88% for LGBM. These methods also achieved accuracy for symptoms of > 85% in all six approaches, with GBM and LGBM showing 90% accuracy. Accuracy matrices, including precision, recall or sensitivity, f1 score, area under the ROC curve (AUC), and log loss values, are shown in Supplementary Table 1 for symptoms data and in Supplementary Table 2 for comorbidity data. The coefficient values for the features (symptoms) are given in Supplementary Table 3, and the features (comorbidities) are given in Supplementary Table 4. Our results indicate that age is the most significant predictor of mortality as well as the gender. We compared both results (most significant features) for symptoms and comorbidities found from different algorithms and got similar predictions. In figure 2 we represent the significance level for symptoms and diseases. After calculating the coefficient values for every algorithm, we measured the symptoms and diseases in the same scale by quantile normalization and using the average normalized values in Figure 2. The most significant symptoms were pneumonia, acute respiratory distress syndrome (ARDS), dyspnea, fever and cough (Supplementary Table 3) and the most significant comorbidities found were hypertension, diabetes and metabolic diseases, chronic kidney disease, cardiovascular disease, chronic obstructive pulmonary disease (COPD), asthma and malignancy in this cohort (Supplementary Table 4).

## Significant pairs of interacting comorbidities and symptoms associated with death in COVID-19

One of the unique findings of this study is the identification of significant pairs of comorbidities and symptoms that are associated with death among COVID-19 patients. For identification of symptom-comorbidity interactions, we applied the Fisher's exact testing procedure. The negative logarithm of the p-values obtained from the tests are presented in Figure-3. We observed that the symptom-comorbidity combination of Pneumonia-Hypertension, Pneumonia-Diabetes and ARDS-Hypertension had the most significant effects on mortality in COVID-19 patients (Figure3).

Taken together, these data provide a comprehensive analysis of the current published literature, as well as a novel machine learning classification analysis using recently aggregated data to identify significant comorbidities and symptom relationships relating to death from COVID-19 disease.

## Discussion

The recent and continuing spread of SARS-CoV-2 has vastly outpaced the ability of many public health care systems around the world to respond and manage. There are many examples from even advanced economies where medical professionals have had to make distressing decisions about prioritization of insufficient care resources. This highlights the critical need for fast and accurate classification of those patients most at risk of severe disease or fatality to best allocate hospital resources during times of crisis.

To this end, we have performed a number of analyses to assess how disease outcome is related to a range of patient comorbidities and clinical features. Firstly, we investigated published COVID-19 clinical data using a conventional meta-analysis. We found almost no evidence of publication bias in this data, and little grey literature sources of use to our study. This may reflect the current strong imperative to rapidly publish any available studies. Our meta-analysis identified COPD, CEVD, CVD, diabetes, malignancy, and hypertension as most significantly associated with COVID-19 severity in the current published literature.

We also obtained and analyzed aggregated COVID-19 patient data (not derived from published clinical trials or retrospective studies) using statistical and machine learning methods. We found that patients most at risk of dying from COVID-19 had particular comorbidities and patient features, most of which were seen in our meta-analysis. Our machine learning analysis of this patient dataset for the classification of deceased versus recovered COVID-19 patients identified COPD, CVD, CKD, diabetes, malignancy, hypertension, and asthma as most significant. These results provide detailed insights into the strength of the relationship between these factors and patients' risk of dying from COVID-19, identifying prognostic factors by largely independent means. This may lead to identification of disease mechanisms of interest by considering pathways that may be common to these comorbidities. Already such considerations have been made with several studies reporting strong evidence for a link between SARS-Cov-2 actions and vascular damage.[41] Further, given that the angiotensin converting enzyme (ACE-2) receptor is used by the virus for entry into host cells, it has been suggested that the already strained ACE-2-Ang-(1-7)-Mas in metabolic disorders may result in a respiratory compromise (42). The role of upregulation of the ACE-2 receptors by ACE inhibitors and angiotensin II receptor blockers used in the management of hypertension, diabetes, and CKD (43) also requires further exploration in elucidating the metabolic pathways that underpin the relationship between these co-morbidities and increased SARS-Cov-2 related severe morbidity and mortality.

It is likely that there are many different factors interacting that lead to the co-incidence of COVID-19 and comorbidities greatly detrimental to patient outcome. We found using machine learning classification methods that age and gender are the most significant predictor of COVID-19 mortality. Indeed, it is likely that in many cohorts, age is strongly associated with the co-occurrence of significant comorbidities as these tend to be age-related diseases. Nevertheless, comorbidities analyzed here such as diabetes, hypertension and asthma do occur across age categories, suggesting mortality in COVID-19 is impacted by other characteristics yet to be identified; perhaps differences in environment and/or genetic predispositions are likely relevant factors for future consideration.

Mechanistically, the association between lung-related comorbidities such as COPD and COVID-19 disease severity are an expected outcome of this study. COPD is a chronic lung condition, often caused by a patient history of smoking. Patients with COPD present with pulmonary damage and chronic breathing difficulty; thus, the co-occurrence of a severe lower respiratory viral infection and pneumonia is a significant challenge, particularly in the elderly. In contrast, the association of severe COVID-19 disease with conditions such as vascular diseases (CVD, CEVD) and diabetes, is perhaps more complex. Data are emerging however that suggests SARS-CoV-2 infection is associated with a severe inflammatory storm that can result in vascular inflammation, as well as myocarditis. Thus cardio-vascular and metabolic diseases are likely compounding the impact of COVID-19; perhaps presenting a therapeutic opportunity for broad-spectrum anti-inflammatory medications, although the data on efficacy remain to be acquired.

An important consideration remains the limitations of the available data for predictive analyses. COVID-19 remains a relatively recent phenomenon, and thus the data may contain biases that cannot as yet be circumvented. For example, the majority of data coming from mainland China presents biases related to population genetics as well as environmental effects that will not be observed in similar

European datasets. Nevertheless, our analysis of this cohort data from 1143 patients comes from repository data acquired from across 141 countries; thus, systematic biases of this kind should be minimal. Additionally however, there may be unidentified reporting biases in global hospital data due to severe under-resourcing and staff shortages in some locations, necessitating priority reporting. Over the coming months, more data will become available from more diverse nations and population groups that will enable fuller investigation of these issues.

## Conclusion

In summary, we have performed a comprehensive meta-analysis of available published literature, as well as a novel machine learning analysis of a separate cohort of COVID-19 patients. We identified significant comorbidities and COVID-19 patient symptoms that are important for consideration when assessing patient needs; something that remains critical at a time where hospitals are often understaffed and under-resourced. Data suggest that the comorbidities most implicated in severe COVID-19 are lung-related, such as COPD and asthma, as well as vascular-related conditions, such as CVD and CEVD. Thus, it is critical that at-risk populations be prioritized in efforts around social isolation and resource allocation during this pandemic. As data continue to be accrued, it will become possible to answer questions regarding gender and age-related comorbidity relationships including medication history as well as population genetics and environmental effects that may be relevant to treatment optimization.

## Methods

This study has two parts - i) meta-analysis of previously published literature, and

ii) machine learning algorithm based analysis on patient-level cohort data.

### Meta-analysis of published data

### Search strategy and study selection

The meta-analysis was conducted according to PRISMA (Preferred Reporting Items for Systematic Reviews and Meta-analysis) and MOOSE (Meta-analysis of Observational Studies in Epidemiology) guidelines.[8-10] Potential and relevant studies were extracted by conducting a systematic search of databases; from January 1, 2019, to April 20, 2020, in PubMed (Medline), Springer, Web of Science, EMBASE, and Cochrane Library databases. This study used keywords for database screening; '2019-nCoV', '2019 novel coronavirus', 'COVID-19', 'clinical characteristics and symptoms of coronavirus'. Databases using comorbidity combinations for all comorbidities studied were also searched, with the following structure: "COVID-19 and diabetes", "COVID-19 and hypertension", "COVID-19 and COPD" and related terms. The list of cited references from selected articles were manually screened to identify missing studies. All articles selected for the meta-analysis were written in English and all search procedures were independently performed by two investigators (MMA and AT).

For this study, articles that described the clinical characteristics of COVID-19 patients were included, particularly symptoms and comorbidities, along with their prevalence and specific information on the distribution of patients on the basis of severity. Key exclusion criteria were: (a) duplicate publications,

(b) case reports, reviews, editorials, letters, or (c) studies that failed to provide sufficient information on clinical patient characteristics, as judged by the two investigators.

### Data extraction for statistical analysis

The two investigators who performed the literature screening also extracted the data independently from the selected studies. Differences in the chosen literature were reconciled by discussion and screening by a third investigator (MAM). We extracted the following variables: first author name, year of publication, number of patients, age, sex, number of patients suffering severe diseases (note that patients were not stratified based on the degree of comorbidity severity or symptom severity), number of non-severe patients where these were reported, patient survival, patients needing intensive care unit (ICU) support, and the prevalence of multiple symptoms and comorbidities. The definition of 'severe' was clearly described in some articles, however not all. We maintained the case definitions as defined by the original authors. The odds ratios (OR) were calculated to describe the severity of clinical symptoms in severe patients compared to non-severe patients. The degree of variability across studies (heterogeneity) was assessed by $I^2$ and Cochran's Q test[11]. Due to the existence of heterogeneity in studies, random-effects models were utilized to estimate the average effect of variables, along with their precision which can provide a more accurate estimate of the 95% confidence intervals (CI).

### Statistical analysis and machine learning analysis of novel aggregate clinical data

### Data collection

We obtained publicly available anonymized clinical data that was derived from both non-hospitalized and hospitalized COVID-19 positive patients; patient diagnoses were based on WHO guidelines.[12] The cases were captured between February 14, 2020, to April 31, 2020. Real-time data was collected from open-source COVID-19 data repositories.[13,14] The data obtained came from a total of 4,81,289 individual patient clinical records from 141 countries.

Summary descriptive statistics for this clinical data are shown in Table 5. The clinical attributes collected included clinical symptoms and signs, details of any comorbidities, date of admission in the hospital, date of confirmation of COVID-19 caseness, date of death or hospital release, details of other associated disease outcomes, as well as demographic data; the latter included age, gender, travel history, and location (e.g., city, province, and country) of the patient. From these data, we filtered for select criteria e.g. patients who are deceased and recovered and released from hospitals. We also excluded patients where data relating to their mortality or recovery from infection was not included. The final filtered dataset included 1,143 COVID-19 patients with detailed clinical information, of whom 319 were reported as deceased and 824 as recovered.

### Selection of significant variables

The focus of this study was to analyze the mortality and survival rates in our filtered 1,143 patient datasets and to relate these rates to comorbidity incidences. We, thus considered respondent age (continuous), sex (male, female), travel history, and the commonly occurring comorbidities, both individually and occurring in multiples. The comorbidities studied included cardiovascular disease (CVD), chronic obstructive pulmonary disease (COPD), cerebrovascular disease (CEVD), chronic kidney disease (CKD), chronic lung disease (CLD), neurodegenerative disease, hypertension, diabetes (type 2), malignancies, infectious diseases, surgical history, asthma, and liver disease. Additionally, we included several clinical symptoms for analysis, including the incidence of fever, cough, pneumonia, acute respiratory distress symptoms (ARDS), dyspnea, fatigue, septic shock, headache, myalgia, diarrhea, and nausea, in order to predict at an early stage and to identify the relationship of the severity or death. We assessed the influence of these variables on the probability of returning a positive diagnosis of SARS-CoV-2 infection.

**Statistical analysis**

Continuous variables were summarized by median along with interquartile range (IQR), and compared by utilizing the Mann-Whitney U test.[15] The frequency of categorical variables was presented in percent and compared with a chi-square test.[16] Moreover, Fisher's exact test[17] was applied to low-frequency cells. A two-sided $\alpha$ (type-I error) less than 0.05 was considered as a measure of statistical significance. All statistical analysis was performed in the R statistical computing environment (version 3.6.1).

**Machine learning algorithms**

In this study, six supervised machine learning algorithms were applied to identify the minimum number of symptoms and comorbidities that were predictive of COVID-19 infection. These algorithms included Random Forest, Decision Tree, Gradient Boosting Machine (GBM), XGBoost (XGB), Support Vector Machine (SVM) and Light Gradient Boosting Machine (LGBM). We extracted the required variables from the raw data, and then performed data cleaning and scaling to pre-process the collected data. Imputation techniques were used to address the missing (2.2%) age and gender values, in particular, the missing age was imputed using random values selected from the age IQR, and gender was imputed randomly according to male and female ratios present in the full dataset. Data was randomly split into training (80% individuals) and testing (20% individuals) data sets to perform machine learning prediction and validation. To measure accuracy, several measures such as precision, recall or sensitivity, f1 score, area under the receiver operating characteristic (ROC) curve (AUC), and log loss values were employed. After achieving high accuracy with the model training, we extracted the features with the highest impact on symptoms and comorbidities classifying a positive COVID-19 infection.

**Author Contributions**

The research presented in this article is a combined effort of twelve authors. The article is prepared by executing several phases of the research. At the early stage of the research, **Sakifa Aktar (SA), Ashis Talukder (AT), Md. Martuza Ahamad (MMA)** had collected the data from several recognized repository on COVID-19 patients according to the direction of **Mohammad Ali Moni (MAM).** Next, the four authors joining with A. H. M. Kamal (AHMK) designed the architecture of the workflow. The other authors are **Jahidur Rahman Khan (JRK), Md. Protikuzzaman (MPK), Nasif Hossain (NH), Julian M.W. Quinn (JMWQ), Mathew A. Summers (MAS), Teng Liaw (TL)**, **Valsamma Eapen (VE)**. Based on that workflow, the authors contributed as follows:
**SA, AT and MMA**: They collected COVID-19 patient's data from the dataset and make them fit for our experiments. They conducted most of the experiments and joined every meeting of the research discussion. They took part in writing the primary draft of the article as well as group-wise reviewing the article at Google drive.

**AHMK:** He partially guided the work along with **MAM**. Consequently, he had to participate in every meeting of the research. He actively took part in writing the primary draft and review phase of the article.
**JRK, MPK, NH, JMWQ, MAS, TL and VE:** They were involved in the writing and reviewing the whole article.
**MAM:** He supervised the whole work. Additionally, he conducted several experiments and took part in every writing phase of the article.
.

## Codes and Data Availability

All the programming codes are available on a GitHub repository and data are available in another GitHub repository [14] and a spreadsheet [13]:

1. https://github.com/m-moni/COVID-19

2. https://github.com/beoutbreakprepared/nCoV2019 (last access April 31, 2020)
3. https://docs.google.com/spreadsheets/d/1Gb5cyg0fjUtsqh3hl_LC5A23zIOXmWH5veBklfSHzg/edit#gid=447265963.

## Competing interests

We do not have any competing interest.

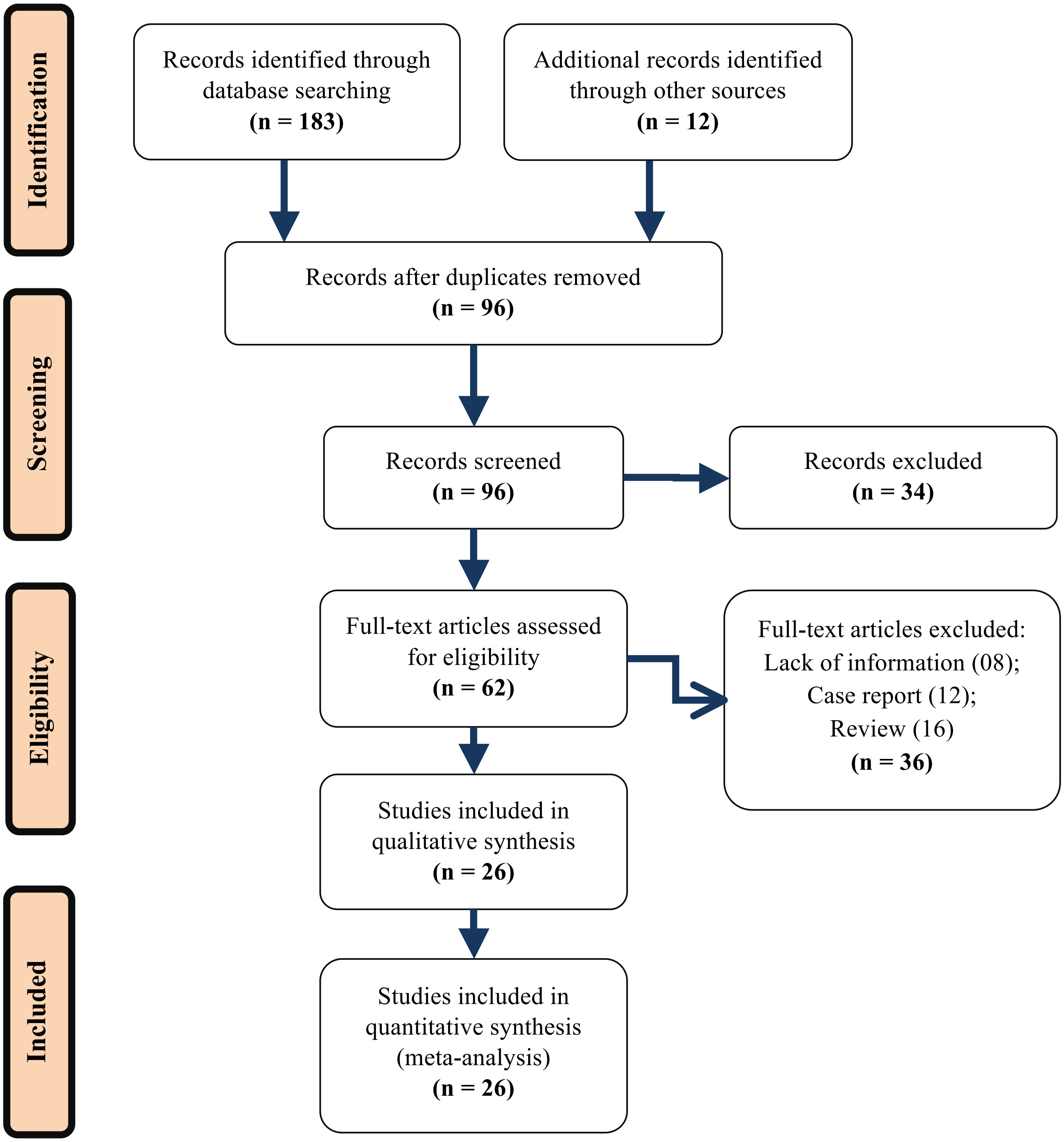

**Figure 1: Flow diagram of literature search for including studies in meta-analysis**

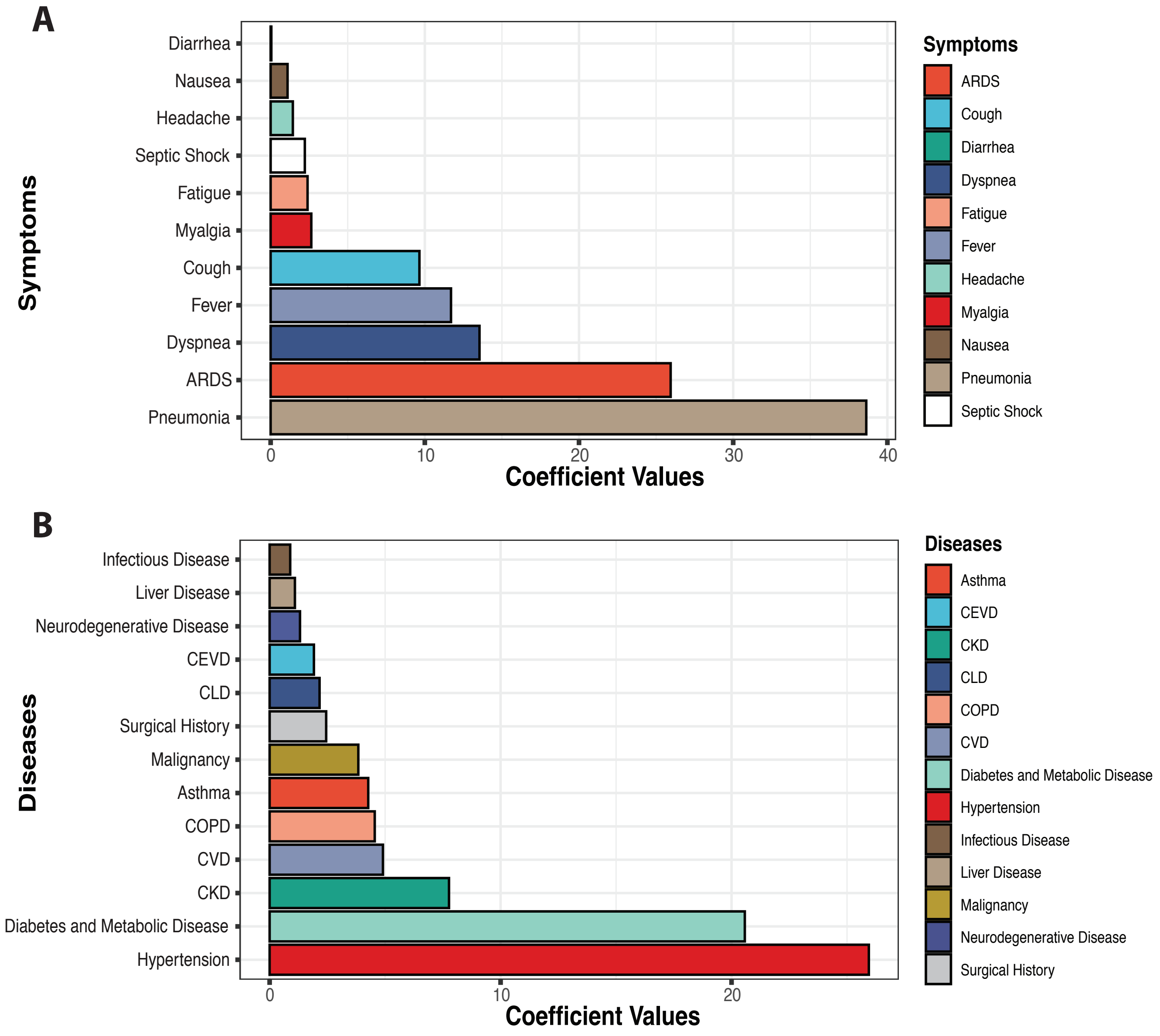

**Figure 2: Machine learning models predict the important symptoms and comorbidities that are associated with the severity or death of COVID-19 patients. The high coefficient values of ML model outcomes mean the higher significant association of death. Figure 2A represents the significance of symptoms that are linked with death. Figure 2B represents the significance of disease comorbidities that are linked with death.**

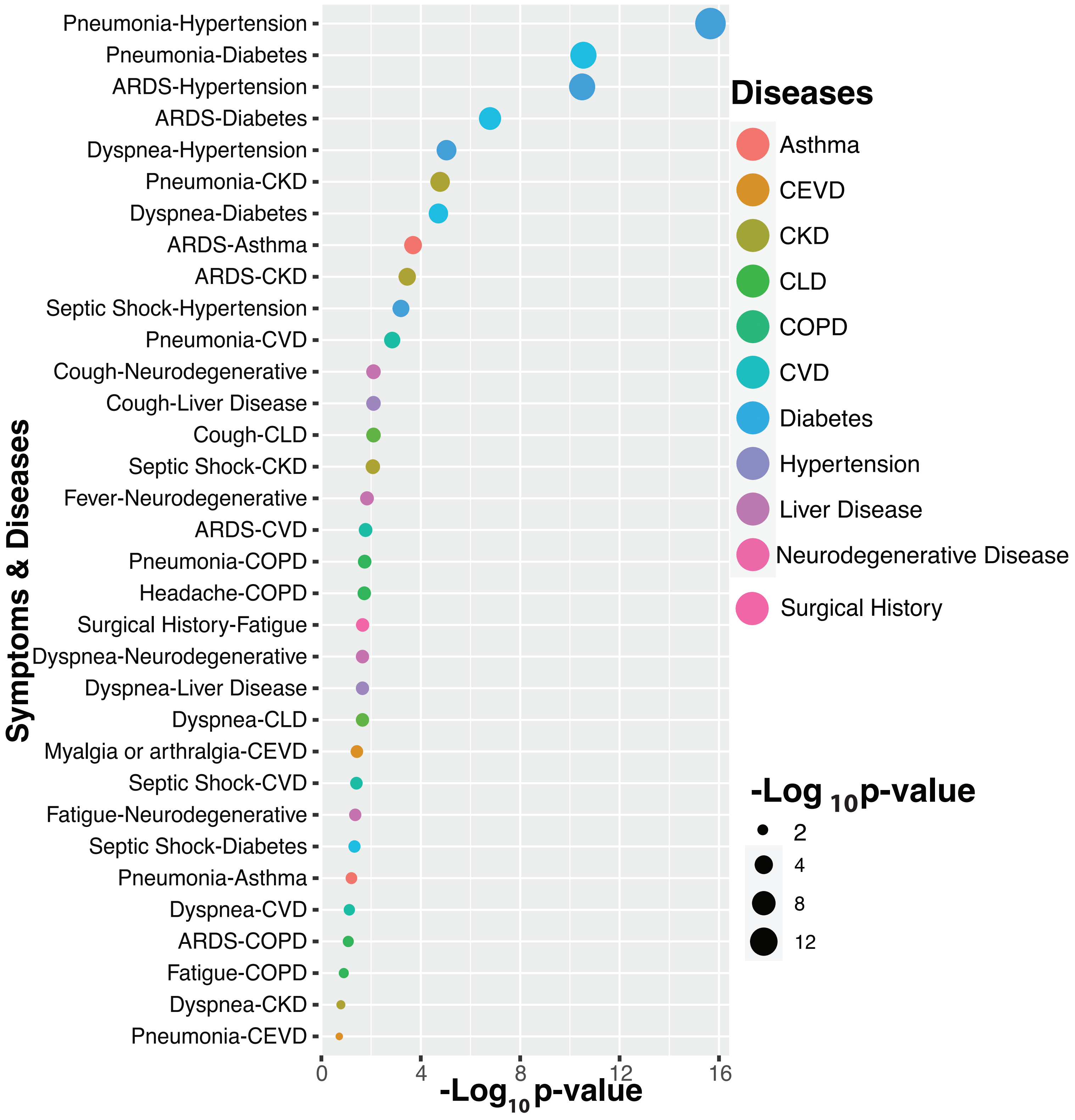

**Figure 3: Association and impact of combined symptoms and comorbidity interactions in COVID-19 deceased patients.**

**Table 1: Summary of study characteristics reported in the selected publications**

| First Author | Year of publication | Country | Sample size (n) | Gender | | Mean/Median age (years) | Severe or death patients n (%) | Reference |
|---|---|---|---|---|---|---|---|---|
| | | | | Male n (%) | Female n (%) | | | |
| Wang et al. | 2020 | China | 138 | 75 (54.35) | 63 (45.65) | 56 | 36 (26.09) | 18 |
| Richardson et al. | 2020 | USA (New York) | 5,700 | 3437 (60.30) | 2263 (39.70) | 63 | 373 (6.54) | 4 |
| Xu et al. | 2020 | China | 62 | 35 (56.45) | 27 (43.55) | 41 | NR | 5 |
| Guan et al. | 2020 | China | 1,099 | 640 (58.23) | 459 (41.77) | 47 | 173 (15.74) | 6 |
| Guan WJ et al. | 2020 | China | 1,590 | 904 (56.86) | 686 (43.14) | 48.9 | 254 (15.97) | 19 |
| Huang et al. | 2020 | China | 41 | 30 (73.17) | 11 (26.83) | 49 | 13 (31.71) | 20 |
| Guo et al. | 2020 | China | 187 | 91 (48.66) | 96 (51.34) | 58.50 | NR | 21 |
| Zhou et al. | 2020 | China | 191 | 119 (62.30) | 72 (37.70) | 56.0 | 66 (34.55) | 22 |
| Zhang et al. | 2020 | China | 140 | 71 (50.71) | 69 (49.29) | 57 | 58 (41.43) | 23 |
| Wu et al. | 2020 | China | 80 | 39 (48.75) | 41 (51.25) | 46.10 | NR | 24 |
| Liu et al. | 2020 | China | 137 | 61 (44.53) | 76 (54.47) | 57 | NR | 25 |
| Liu J et al. | 2020 | China | 61 | 31 (50.82) | 30 (49.18) | 40 | 17 (27.87) | 26 |
| Chen et al. | 2020 | China | 99 | 67 (67.68) | 32 (32.32) | 55.5 | NR | 27 |
| Yang et al. | 2020 | China | 52 | 35 (67.31) | 17 (32.69) | 59.7 | 52 (100.00) | 28 |
| Wu C et al. | 2020 | China | 201 | 128 (63.68) | 73 (36.32) | 51 | 53 (26.37) | 29 |
| Jie Li et al. | 2020 | China | 17 | 9 (52.94) | 8 (47.06) | 45.1 | NR | 30 |
| Liu W et al. | 2020 | China | 78 | 39 (50.00) | 39 (50.00) | 38 | NR | 31 |
| Mo etal. | 2020 | China | 155 | 86 (55.48) | 69 (44.52) | 54 | 55 (35.48) | 32 |
| Du et al. | 2020 | China | 85 | 62 (72.94) | 23 (27.06) | 65.8 | NR | 33 |
| Rong-Hui et al. | 2020 | China | 179 | 97 (54.19) | 82 (45.81) | 57.6 | NR | 34 |
| Feng et al. | 2020 | China | 476 | 271 (56.93) | 205 (43.07) | 53 | 26 (5.46) | 35 |
| Chen et al. | 2020 | China | 274 | 171 (62.41) | 103 (37.59) | 62 | 113 (41.24) | 36 |
| Grasselli et al. | 2020 | Italy | 1,591 | 1304 (81.96) | 287 (18.04) | 63 | 1,591 (100.00) | 37 |
| Deng et al. | 2020 | China | 225 | 73 (32.44) | 152 (67.56) | 69 | NR | 38 |
| Wang et al. | 2020 | China | 339 | 166 (48.97) | 173 (51.03) | 69 | 65 (19.17) | 39 |
| Chen TL et al. | 2020 | China | 203 | 108 (53.20) | 95 (76.80) | 54 | 19 (9.36) | 40 |
| **Total** | - | - | 13,400 | 8149 (60.81) | 5206 (39.19) | - | 2964 (22.11%) | - |

**NR=Not Reported**

**Table 2: Prevalence of symptoms in COVID-19 patients in the selected studies**

| First Author | Year of publication | Sample size (n) | Clinical Symptoms | | | | | | | | | Reference |
|---|---|---|---|---|---|---|---|---|---|---|---|---|
| | | | Fever (%) | Cough (%) | Fatigue (%) | Anorexia (%) | Myalgia (%) | Dyspnea (%) | Diarrhea (%) | Nausea (%) | Headache (%) | |
| Wang et al. | 2020 | 138 | 98.55 | 59.42 | 69.57 | 39.86 | 34.78 | 31.16 | 26.09 | 10.14 | 6.52 | 18 |
| Richardson et al. | 2020 | 5,700 | NR | NR | NR | NR | NR | NR | 6.54 | NR | NR | 4 |
| Xu et al. | 2020 | 62 | 77.42 | 80.65 | 51.61 | NR | 51.61 | NR | NR | NR | 33.87 | 5 |
| Guan et al. | 2020 | 1,099 | 43.04 | 67.79 | 38.13 | NR | NR | NR | 15.74 | 5.00 | 13.65 | 6 |
| Guan WJ et al. | 2020 | 1,590 | 84.97 | 66.16 | 36.73 | NR | NR | NR | 15.97 | 5.03 | 12.89 | 19 |
| Huang et al. | 2020 | 41 | 97.56 | 75.61 | 43.90 | NR | 43.90 | 53.66 | 31.71 | NR | 7.32 | 20 |
| Guo et al. | 2020 | 187 | NR | NR | NR | NR | NR | NR | NR | NR | NR | 21 |
| Zhou et al. | 2020 | 191 | 94.24 | 79.06 | 23.04 | NR | 15.18 | NR | 34.55 | 3.66 | NR | 22 |
| Zhang et al. | 2020 | 140 | 78.57 | 64.29 | 64.29 | 12.14 | NR | NR | 41.43 | 17.14 | NR | 23 |
| Wu et al. | 2020 | 80 | NR | NR | NR | NR | NR | NR | NR | NR | NR | 24 |
| Liu et al. | 2020 | 137 | 81.75 | 48.18 | 32.12 | NR | 32.12 | 18.98 | NR | 62.04 | 9.49 | 25 |
| Liu J et al. | 2020 | 61 | 98.36 | 63.93 | 57.38 | NR | NR | 4.92 | 27.87 | 8.20 | 34.43 | 26 |
| Chen et al. | 2020 | 99 | 82.83 | 81.82 | NR | NR | NR | NR | NR | 1.01 | 8.08 | 27 |
| Yang et al. | 2020 | 52 | 98.08 | 28.85 | NR | NR | 3.85 | 23.08 | 100.00 | NR | 1.92 | 28 |
| Wu C et al. | 2020 | 201 | 93.53 | 81.09 | 32.34 | NR | 32.34 | 39.80 | 26.37 | NR | NR | 29 |
| Jie Li et al. | 2020 | 17 | 70.59 | 76.47 | 47.06 | NR | 23.53 | NR | NR | NR | NR | 30 |
| Liu W et al. | 2020 | 78 | NR | 43.59 | NR | NR | NR | NR | NR | NR | NR | 31 |
| Mo etal. | 2020 | 155 | 81.29 | 62.58 | 38.71 | 16.77 | NR | 1.29 | 35.48 | 1.94 | 5.16 | 32 |
| Du et al. | 2020 | 85 | 91.76 | NR | 58.82 | 56.47 | 16.47 | 70.59 | NR | NR | 4.71 | 33 |
| Rong-Hui et al. | 2020 | 179 | 98.88 | 81.56 | 39.66 | NR | 18.99 | 49.72 | NR | NR | 9.50 | 34 |
| Feng et al. | 2020 | 476 | 81.93 | NR | 56.51 | NR | 11.55 | NR | 5.46 | NR | NR | 35 |
| Chen et al. | 2020 | 274 | 90.88 | 67.52 | 50.00 | 24.09 | 21.90 | NR | 41.24 | 8.76 | 11.31 | 36 |
| Grasselli et al. | 2020 | 1,591 | NR | NR | NR | NR | NR | NR | 100.00 | NR | NR | 37 |
| Deng et al. | 2020 | 225 | 42.22 | 20.89 | 13.33 | NR | 13.33 | 34.22 | NR | NR | NR | 38 |
| Wang et al. | 2020 | 339 | 91.74 | 52.80 | 39.82 | 27.73 | 4.72 | 40.71 | 19.17 | 3.83 | 3.54 | 39 |
| Chen TL et al. | 2020 | 203 | 89.16 | 60.10 | 7.88 | 2.96 | 26.60 | 1.48 | 9.36 | 1.48 | 4.93 | 40 |
| **Overall prevalence (95% CI)** | | | 88.26 (81.31, 92.84) | 63.68 (57.49, 69.45) | 40.48 (34.49, 48.77) | 21.92 (13.50, 33.56 ) | 21.01 (15.50, 27.82) | 26.49 (18.50, 36.39) | 7.60 (4.89, 11.63) | 6.50 (3.10, 13.10) | 9.84 (7.38, 13.00) | - |
| $I^2$ (%) | | | 98 | 94 | 94 | 94 | 92 | 93 | 93 | 97 | 87 | - |
| ***p* for heterogeneity** | | | <0.01 | <0.01 | <0.01 | <0.01 | <0.01 | <0.01 | <0.01 | <0.01 | <0.01 | - |

**Note: Meta-analysis for the prevalence was calculated from random-effects model analysis (see, Supplementary Figure 1 for details); NR=Not Reported**

**Table 3: Prevalence of comorbidities in COVID-19 patients in the selected studies**

| First Author | Year of publication | Sample size (n) | Comorbidities | | | | | | | | | Reference |
|---|---|---|---|---|---|---|---|---|---|---|---|---|
| | | | Hypertension (%) | Diabetes (%) | CVD (%) | Malignancy (%) | COPD (%) | CEVD (%) | CKD (%) | CLD (%) | Smoking (%) | |
| Wang et al. | 2020 | 138 | 31.16 | 10.14 | 14.49 | 7.25 | 2.90 | 5.07 | 2.90 | 2.90 | NR | 18 |
| Richardson et al. | 2020 | 5,700 | 53.09 | 31.72 | 14.46 | 5.61 | 5.04 | NR | 7.95 | 0.19 | 47.21 | 4 |
| Xu et al. | 2020 | 62 | 8.06 | 1.61 | NR | NR | 1.61 | 1.61 | 1.61 | 11.29 | NR | 5 |
| Guan et al. | 2020 | 1,099 | 15.01 | 7.37 | 2.46 | 0.91 | 1.09 | 1.36 | 0.73 | 2.09 | 14.37 | 6 |
| Guan WJ et al. | 2020 | 1,590 | 16.92 | 8.18 | 3.71 | 8.18 | 1.51 | 1.89 | 16.92 | 1.51 | 6.98 | 19 |
| Huang et al. | 2020 | 41 | 14.63 | 19.51 | 4.88 | 2.44 | 2.44 | NR | NR | 2.44 | 7.31 | 20 |
| Guo et al. | 2020 | 187 | 32.62 | 14.97 | 11.23 | NR | 2.14 | NR | 3.21 | NR | 9.62 | 21 |
| Zhou et al. | 2020 | 191 | 30.37 | 18.85 | 7.85 | NR | NR | NR | 1.05 | NR | 5.75 | 22 |
| Zhang et al. | 2020 | 140 | 30.00 | 12.14 | 7.14 | NR | 1.43 | NR | 1.43 | 5.71 | 6.42 | 23 |
| Wu et al. | 2020 | 80 | NR | NR | 31.25 | 5.00 | NR | NR | 1.25 | 1.25 | NR | 24 |
| Liu et al. | 2020 | 137 | 9.49 | 10.22 | 7.30 | 1.46 | 1.46 | NR | NR | NR | NR | 25 |
| Liu J et al. | 2020 | 61 | 19.67 | 8.20 | NR | NR | 8.20 | 1.64 | NR | NR | 6.55 | 26 |
| Chen et al. | 2020 | 99 | NR | 12.12 | 40.40 | NR | 1.01 | NR | NR | NR | NR | 27 |
| Yang et al. | 2020 | 52 | NR | 3.85 | 7.69 | 1.92 | NR | NR | NR | NR | 3.84 | 28 |
| Wu C et al. | 2020 | 201 | 19.40 | 10.95 | 3.98 | NR | 2.49 | NR | 1.00 | 3.48 | NR | 29 |
| Jie Li et al. | 2020 | 17 | 5.88 | NR | NR | NR | NR | NR | NR | NR | 17.64 | 30 |
| Liu W et al. | 2020 | 78 | 10.26 | 6.41 | NR | 5.13 | 2.56 | NR | NR | NR | 6.41 | 31 |
| Mo etal. | 2020 | 155 | 23.87 | 9.68 | 9.68 | 4.52 | 3.23 | 4.52 | 3.87 | 4.52 | 3.87 | 32 |
| Du et al. | 2020 | 85 | 37.65 | 22.35 | 11.76 | 7.06 | 2.35 | 8.24 | 3.53 | 5.88 | NR | 33 |
| Rong-Hui et al. | 2020 | 179 | 32.40 | 18.44 | 16.20 | 2.23 | NR | NR | 2.23 | NR | NR | 34 |
| Feng et al. | 2020 | 476 | NR | NR | NR | NR | NR | NR | NR | NR | 9.24 | 35 |
| Chen et al. | 2020 | 274 | 33.94 | 17.15 | 8.76 | 2.55 | 6.57 | 1.46 | 1.46 | 4.01 | 6.93 | 36 |
| Grasselli et al. | 2020 | 1,591 | 31.99 | 11.31 | 14.02 | 5.09 | 2.64 | NR | 2.26 | 1.76 | NR | 37 |
| Deng et al. | 2020 | 225 | 17.78 | 7.56 | 5.78 | 2.67 | 9.78 | NR | NR | NR | NR | 38 |
| Wang et al. | 2020 | 339 | 40.71 | 15.93 | 15.63 | 4.42 | 6.19 | 6.19 | 3.83 | 0.59 | NR | 39 |
| Chen TL et al. | 2020 | 203 | 21.18 | 7.88 | 7.88 | 3.45 | 3.94 | 4.43 | 3.94 | 3.94 | NR | 40 |
| **Overall prevalence (95% CI)** | | | 23.41 (17.63, 30.63) | 11.84 (8.27, 18.14) | 10.00 (7.68, 12.93) | 4.09 (3.18, 5.24) | 3.18 (2.33, 4.34) | 3.23 (2.02, 5.13) | 2.78 (1.74, 4.41) | 2.50 (1.51, 4.11) | 8.83 (4.19, 17.69) | - |
| $I^2$ (%) | | | 98 | 97 | 94 | 79 | 82 | 79 | 95 | 88 | 99 | - |
| ***p* for heterogeneity** | | | <0.01 | <0.01 | <0.01 | <0.01 | <0.01 | <0.01 | <0.01 | <0.01 | <0.01 | - |

**CVD= Cardiovascular disease; COPD=Chronic obstructive pulmonary disease; CEVD= Cerebrovascular disease; CKD=Chronic Kidney Disease; CLD=Chronic lung disease. Note: Meta-analysis for the prevalence was calculated from random-effects model analysis (see, Supplementary Figure 1 for details)**

**Table 4: Odds ratio representing the severity of comorbidities and symptoms in COVID-19 patients c from meta-analysis of published data**

| Outcomes | Number of Studies | Number of Patients | Odds ratio (95% CI) | $I^2$ % (P value) |
|---|---|---|---|---|
| **Comorbidities** | - | - | - | - |
| Hypertension | 10 | 2641 | 1.81 (1.49, 2.20) | 0 (0.72) |
| Diabetes | 11 | 2693 | 2.08 (1.39, 3.10) | 46 (0.05) |
| CVD | 6 | 1150 | 3.46 (2.05, 5.87) | 32 (0.21) |
| Malignancy | 6 | 1161 | 2.04 (1.02, 4.07) | 0 (0.67) |
| COPD | 8 | 2176 | 4.76 (2.69, 8.39) | 0 (0.97) |
| CEVD | 6 | 2208 | 4.54 (2.29, 8.99) | 16 (0.31) |
| CKD | 8 | 2539 | 3.22 (1.70, 6.10) | 0 (0.93) |
| Smoking | 6 | 1920 | 1.74 (1.25, 2.42) | 0 (0.88) |
| **Clinical Symptoms** | - | | - | - |
| Fever | 11 | 2693 | 1.04 (0.85, 1.28) | 42 (0.07) |
| Cough | 11 | 2693 | 1.12 (0.91, 1.38) | 41 (0.09) |
| Fatigue | 10 | 2641 | 1.14 (0.96, 1.36) | 0 (0.99) |
| Anorexia | 5 | 1046 | 1.56 (0.93, 2.62) | 62 (0.03) |
| Myalgia | 7 | 1238 | 0.78 (0.54, 1.13) | 0 (0.68) |
| Dyspnea | 7 | 989 | 2.43 (1.52, 3.89) | 19 (0.29) |
| Diarrhea | 9 | 2600 | 1.14 (0.81, 1.61) | 8 (0.37) |
| Nausea | 7 | 2242 | 0.93 (0.58, 1.47) | 15 (0.31) |
| Headache | 6 | 1779 | 1.04 (0.69, 1.56) | 11 (0.34) |

**Note: CVD= Cardiovascular disease; COPD=Chronic obstructive pulmonary disease; CEVD= Cerebrovascular disease; CKD Kidney Disease; CLD= Chronic lung disease; Odds ratio: Meta-Analysis for overall odds ratio (see, Supplementary Figure details); P value of Egger's test: Assessing the publication bias (see, Supplementary Figure-4 details)**

**Table 5: Association between patient survival and selected demographic characteristics, comorbidities and clinical symptoms.**

| Characteristics | All Patients, n=1143 (%) | Patient's Condition | | P value |
|---|---|---|---|---|
| | | Dead, n=319 (%) | Survived, n=824 (%) | |
| Age, median (IQR) | 51 (36-66) | 74 (63-82) | 46 (32-53) | <0.001 |
| Gender | | | | <0.001 |
| Female | 388 (33.95) | 104 (32.60) | 284 (34.47) | - |
| Male | 600 (52.49) | 197 (61.76) | 403 (48.91) | - |
| Unknown | 155 (13.56) | 18 (5.64) | 137 (16.63) | - |
| Travel History | 370 (32.37) | 80 (25.08) | 290 (35.19) | 0.001 |
| **Comorbidities** | | | | |
| CVD | 21 (1.84) | 16 (5.01) | 5 (0.61) | <0.001 |
| CEVD | 4 (0.35) | 4 (1.25) | 0 | 0.005 |
| CLD | 7 (0.61) | 3 (0.94) | 4 (0.49) | 0.406 |
| Malignancy | 9 (0.79) | 4 (1.25) | 5 (0.61) | 0.275 |
| Diabetes and Metabolic Disease | 80 (6.99) | 61 (19.12) | 19 (2.31) | <0.001 |
| Liver Disease | 2 (0.17) | 2 (0.63) | 0 | 0.078 |
| CKD | 20 (1.75) | 13 (4.08) | 7 (0.85) | <0.001 |
| Neurodegenerative Disease | 2 (0.17) | 2 (0.63) | 0 | 0.078 |
| Infectious Disease | 2 (0.17) | 0 | 2 (0.24) | 1.00 |
| Surgical History | 3 (0.26) | 1 (0.31) | 2 (0.24) | 1.00 |
| COPD | 8 (0.69) | 6 (1.88) | 2 (0.24) | 0.007 |
| Asthma | 10 (0.87) | 5 (1.57) | 5 (0.61) | 0.226 |
| Hypertension | 100 (8.74) | 74 (23.19) | 26 (3.15) | <0.001 |
| **Symptoms** | | | | |
| Headache | 11 (0.96) | 1 (0.31) | 10 (1.21) | 0.308 |
| Fever | 145 (12.68) | 39 (12.22) | 106 (12.86) | 0.848 |
| Cough | 113 (9.88) | 29 (9.09) | 84 (10.19) | 0.653 |
| Fatigue | 25 (2.19) | 8 (2.51) | 17 (2.06) | 0.814 |
| Nausea | 3 (0.26) | 1 (0.31) | 2 (0.24) | 1.00 |
| Diarrhea | 7 (0.61) | 1 (0.31) | 6 (0.73) | 0.681 |
| Myalgia | 11 (0.96) | 3 (0.94) | 8 (0.97) | 1.00 |
| Dyspnea | 59 (5.16) | 48 (15.04) | 11 (1.33) | <0.001 |
| Pneumonia | 74 (6.47) | 66 (20.69) | 6 (0.73) | <0.001 |
| ARDS | 67 (5.86) | 60 (18.81) | 7 (0.85) | <0.001 |
| Septic Shock | 18 (1.57) | 16 (5.02) | 2 (0.24) | <0.001 |
| **Comorbidity Number** | | | | <0.001 |
| No Comorbidity | 990 (86.61) | 214 (67.08) | 775 (94.05) | - |
| Comorbidity=1 | 61 (5.34) | 35 (10.97) | 27 (3.28) | - |
| Comorbidity>1 | 90 (7.87) | 70 (21.94) | 5 (0.61) | - |

**Note: CVD= Cardiovascular disease; COPD=Chronic obstructive pulmonary disease; CEVD= Cerebrovascular disease; CKD=Chronic Kidney Disease; CLD= Chronic lung disease; ARDS=Acute Respiratory Distress Syndrome**

## A. Fever

| Study | Fever | Total | Prevalence | 95% C.I. |
|---|---|---|---|---|
| Wang et al. | 136 | 138 | 98.55 | [94.39; 99.64] |
| Xu et al. | 48 | 62 | 77.42 | [65.40; 86.15] |
| Guan et al. | 473 | 1099 | 43.04 | [40.14; 45.99] |
| Guan WJ et al. | 1351 | 1590 | 84.97 | [83.13; 86.64] |
| Huang et al. | 40 | 41 | 97.56 | [84.61; 99.66] |
| Zhou et al. | 180 | 191 | 94.24 | [89.90; 96.78] |
| Zhang et al. | 110 | 140 | 78.57 | [71.00; 84.59] |
| Liu et al. | 112 | 137 | 81.75 | [74.39; 87.36] |
| Liu et al. | 60 | 61 | 98.36 | [89.27; 99.77] |
| Chen et al. | 82 | 99 | 82.83 | [74.10; 89.05] |
| Yang et al. | 51 | 52 | 98.08 | [87.57; 99.73] |
| Wu et al. | 188 | 201 | 93.53 | [89.18; 96.21] |
| Jie Li et al. | 12 | 17 | 70.59 | [45.81; 87.20] |
| Mo etal. | 126 | 155 | 81.29 | [74.37; 86.68] |
| Du et al. | 78 | 85 | 91.76 | [83.72; 96.02] |
| Rong-Hui et al. | 177 | 179 | 98.88 | [95.64; 99.72] |
| Feng et al. | 390 | 476 | 81.93 | [78.22; 85.14] |
| Chen et al. | 249 | 274 | 90.88 | [86.85; 93.76] |
| Deng et al. | 95 | 225 | 42.22 | [35.93; 48.77] |
| Wang et al. | 311 | 339 | 91.74 | [88.30; 94.24] |
| Chen TL et al. | 181 | 203 | 89.16 | [84.09; 92.76] |
| **Random effects model** | | . | **88.25** | **[81.31; 92.84]** |

Heterogeneity: $I^2 = 98\%$, $\tau^2 = 1.4403$, $\chi^2_{20} = 942.36$ ($p < 0.01$)

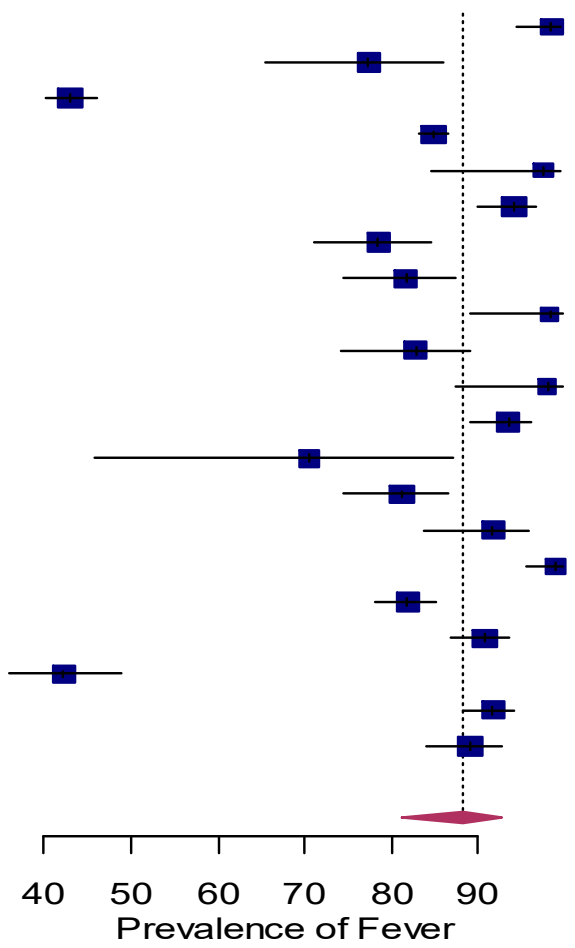

## B. Cough

| Study | Cough | Total | Prevalence | 95% C.I. |
|---|---|---|---|---|
| Wang et al. | 82 | 138 | 59.42 | [51.04; 67.29] |
| Xu et al. | 50 | 62 | 80.65 | [68.93; 88.67] |
| Guan et al. | 745 | 1099 | 67.79 | [64.97; 70.49] |
| Guan WJ et al. | 1052 | 1590 | 66.16 | [63.80; 68.45] |
| Huang et al. | 31 | 41 | 75.61 | [60.32; 86.34] |
| Zhou et al. | 151 | 191 | 79.06 | [72.71; 84.25] |
| Zhang et al. | 90 | 140 | 64.29 | [56.02; 71.78] |
| Liu et al. | 66 | 137 | 48.18 | [39.94; 56.52] |
| Liu j et al. | 39 | 61 | 63.93 | [51.25; 74.93] |
| Chen et al. | 81 | 99 | 81.82 | [72.97; 88.23] |
| Yang et al. | 15 | 52 | 28.85 | [18.20; 42.48] |
| Wu et al. | 163 | 201 | 81.09 | [75.08; 85.93] |
| Jie Li et al. | 13 | 17 | 76.47 | [51.45; 90.88] |
| Liu W et al. | 34 | 78 | 43.59 | [33.06; 54.73] |
| Mo etal. | 97 | 155 | 62.58 | [54.71; 69.84] |
| Rong-Hui et al. | 146 | 179 | 81.56 | [75.20; 86.59] |
| Chen et al. | 185 | 274 | 67.52 | [61.75; 72.80] |
| Deng et al. | 47 | 225 | 20.89 | [16.07; 26.69] |
| Wang et al. | 179 | 339 | 52.80 | [47.48; 58.07] |
| Chen TL et al. | 122 | 203 | 60.10 | [53.21; 66.61] |
| **Random effects model** | | . | **63.68** | **[57.49; 69.45]** |

Heterogeneity: $I^2 = 94\%$, $\tau^2 = 0.3029$, $\chi^2_{19} = 305.96$ ($p < 0.01$)

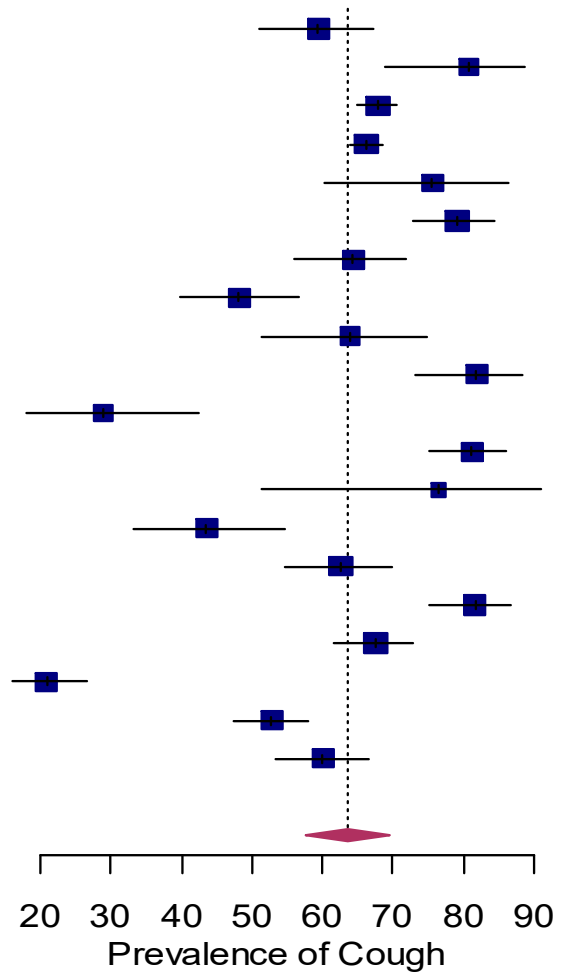

## C. Fatigue

| Study | Fatigue | Total | Prevalence | 95% C.I. |
|---|---|---|---|---|
| Wang et al. | 96 | 138 | 69.57 | [61.40; 76.66] |
| Xu et al. | 32 | 62 | 51.61 | [39.33; 63.71] |
| Guan et al. | 419 | 1099 | 38.13 | [35.30; 41.04] |
| Guan WJ et al. | 584 | 1590 | 36.73 | [34.39; 39.13] |
| Huang et al. | 18 | 41 | 43.90 | [29.69; 59.19] |
| Zhou et al. | 44 | 191 | 23.04 | [17.61; 29.54] |
| Zhang et al. | 90 | 140 | 64.29 | [56.02; 71.78] |
| Liu et al. | 44 | 137 | 32.12 | [24.84; 40.38] |
| Liu j et al. | 35 | 61 | 57.38 | [44.76; 69.10] |
| Wu et al. | 65 | 201 | 32.34 | [26.23; 39.11] |
| Jie Li et al. | 8 | 17 | 47.06 | [25.54; 69.73] |
| Mo etal. | 60 | 155 | 38.71 | [31.37; 46.60] |
| Du et al. | 50 | 85 | 58.82 | [48.12; 68.75] |
| Rong-Hui et al. | 71 | 179 | 39.66 | [32.76; 47.00] |
| Feng et al. | 269 | 476 | 56.51 | [52.02; 60.90] |
| Chen et al. | 137 | 274 | 50.00 | [44.11; 55.89] |
| Deng et al. | 30 | 225 | 13.33 | [9.48; 18.43] |
| Wang et al. | 135 | 339 | 39.82 | [34.74; 45.13] |
| Chen TL et al. | 16 | 203 | 7.88 | [4.88; 12.48] |
| **Random effects model** | | . | **40.48** | **[34.49; 46.77]** |

Heterogeneity: $I^2 = 94\%$, $\tau^2 = 0.2849$, $\chi^2_{18} = 319.55$ ($p < 0.01$)

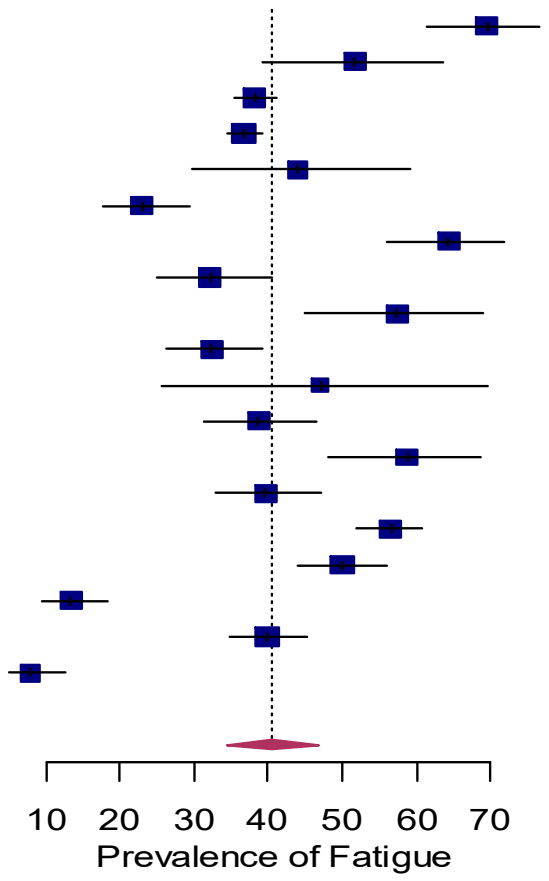

## D. Anorexia

| Study | Anorexia | Total | Prevalence | 95% C.I. |
|---|---|---|---|---|
| Wang et al. | 55 | 138 | 39.86 | [32.03; 48.23] |
| Zhang et al. | 17 | 140 | 12.14 | [7.68; 18.67] |
| Mo etal. | 26 | 155 | 16.77 | [11.68; 23.50] |
| Du et al. | 48 | 85 | 56.47 | [45.80; 66.58] |
| Chen et al. | 66 | 274 | 24.09 | [19.39; 29.50] |
| Wang et al. | 94 | 339 | 27.73 | [23.22; 32.74] |
| Chen TL et al. | 6 | 203 | 2.96 | [1.33; 6.42] |
| **Random effects model** | | . | **21.92** | **[13.50; 33.56]** |

Heterogeneity: $I^2 = 94\%$, $\tau^2 = 0.5751$, $\chi^2_6 = 104.15$ ($p < 0.01$)

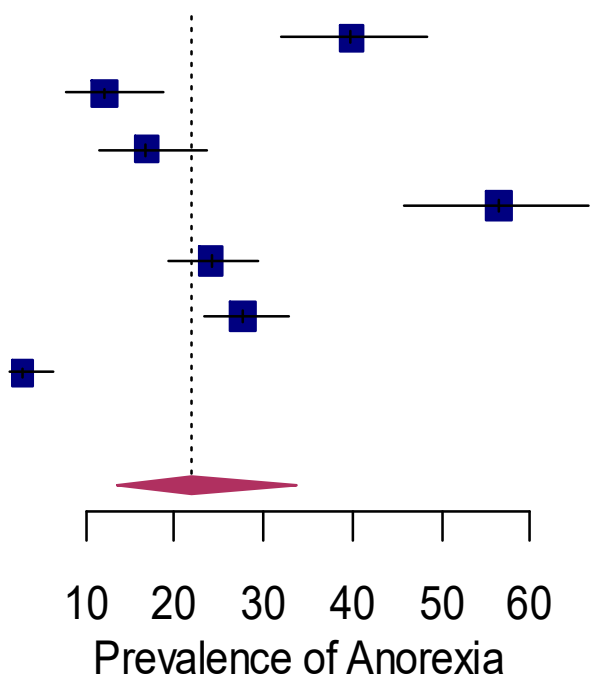

## E. Myalgia

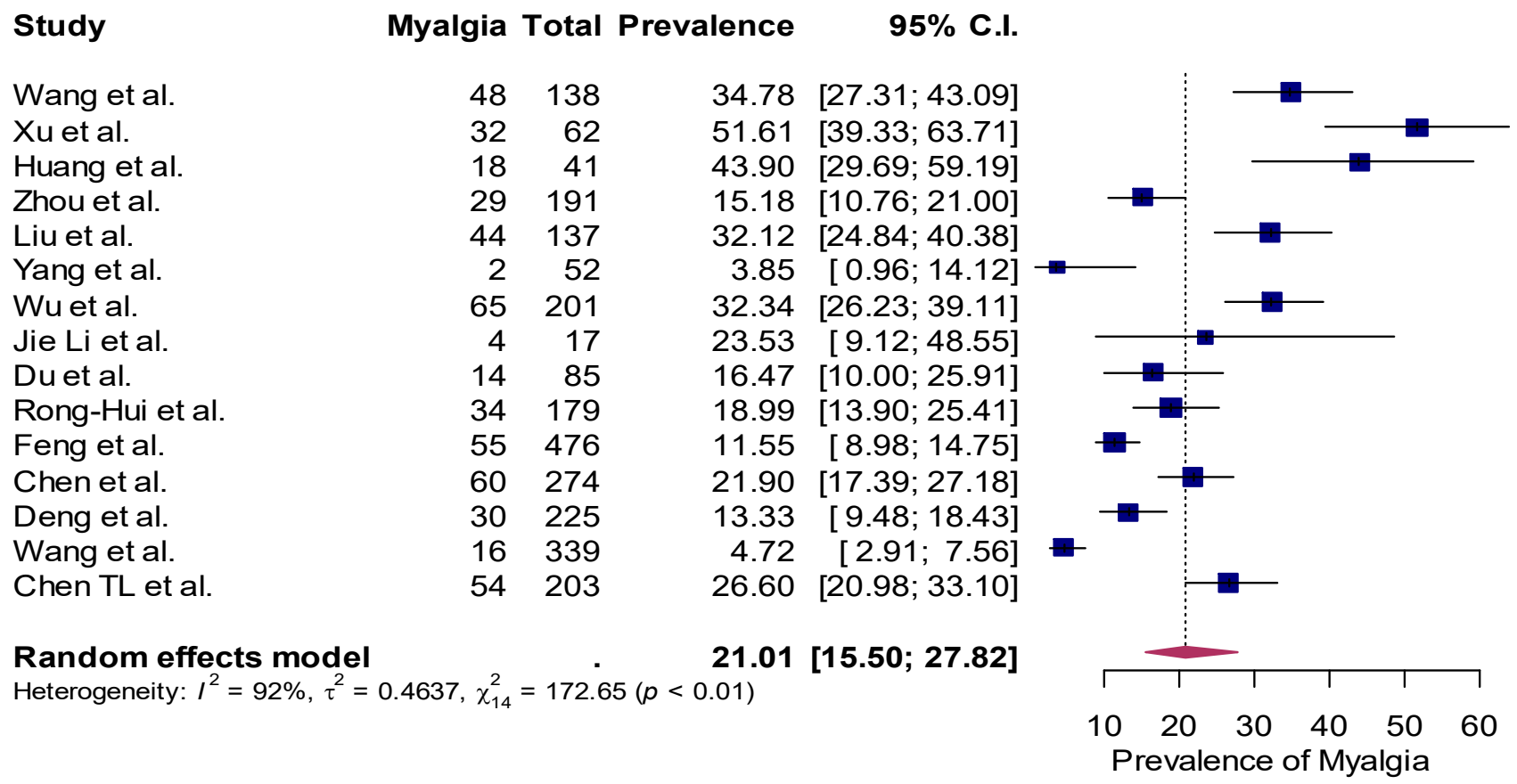

## F. Dyspnea

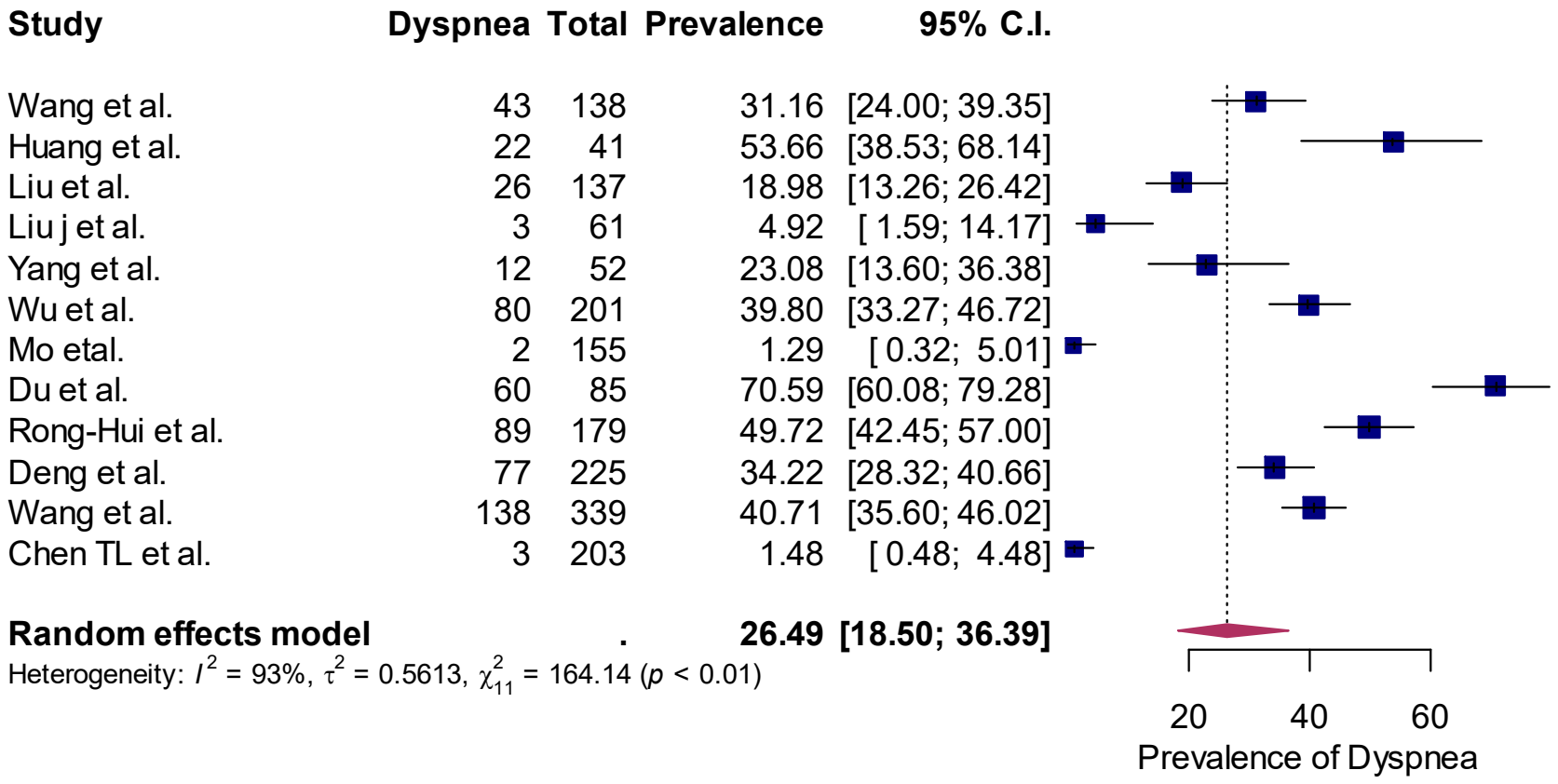

## G. Diarrhea

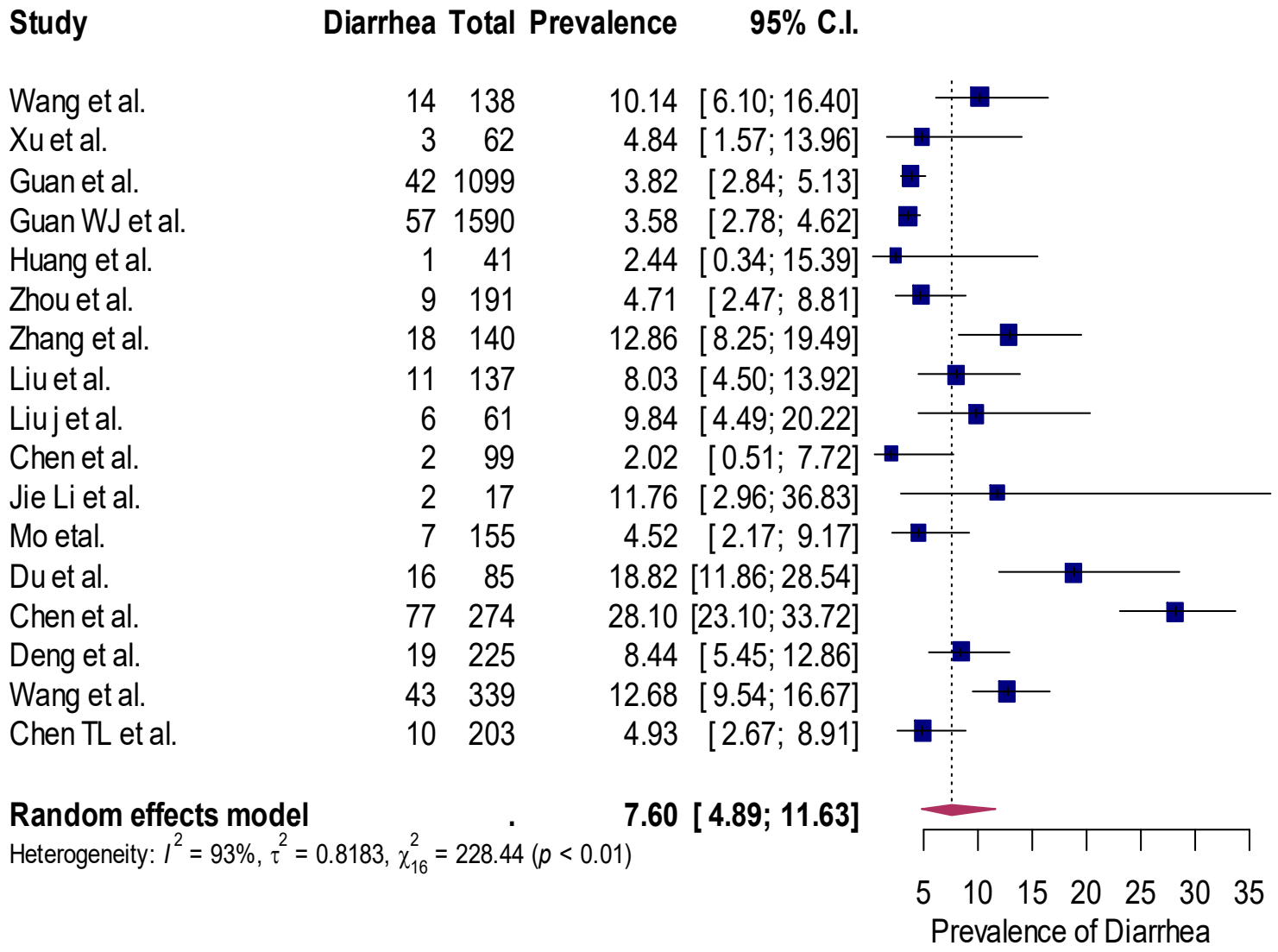

## H. Nausea

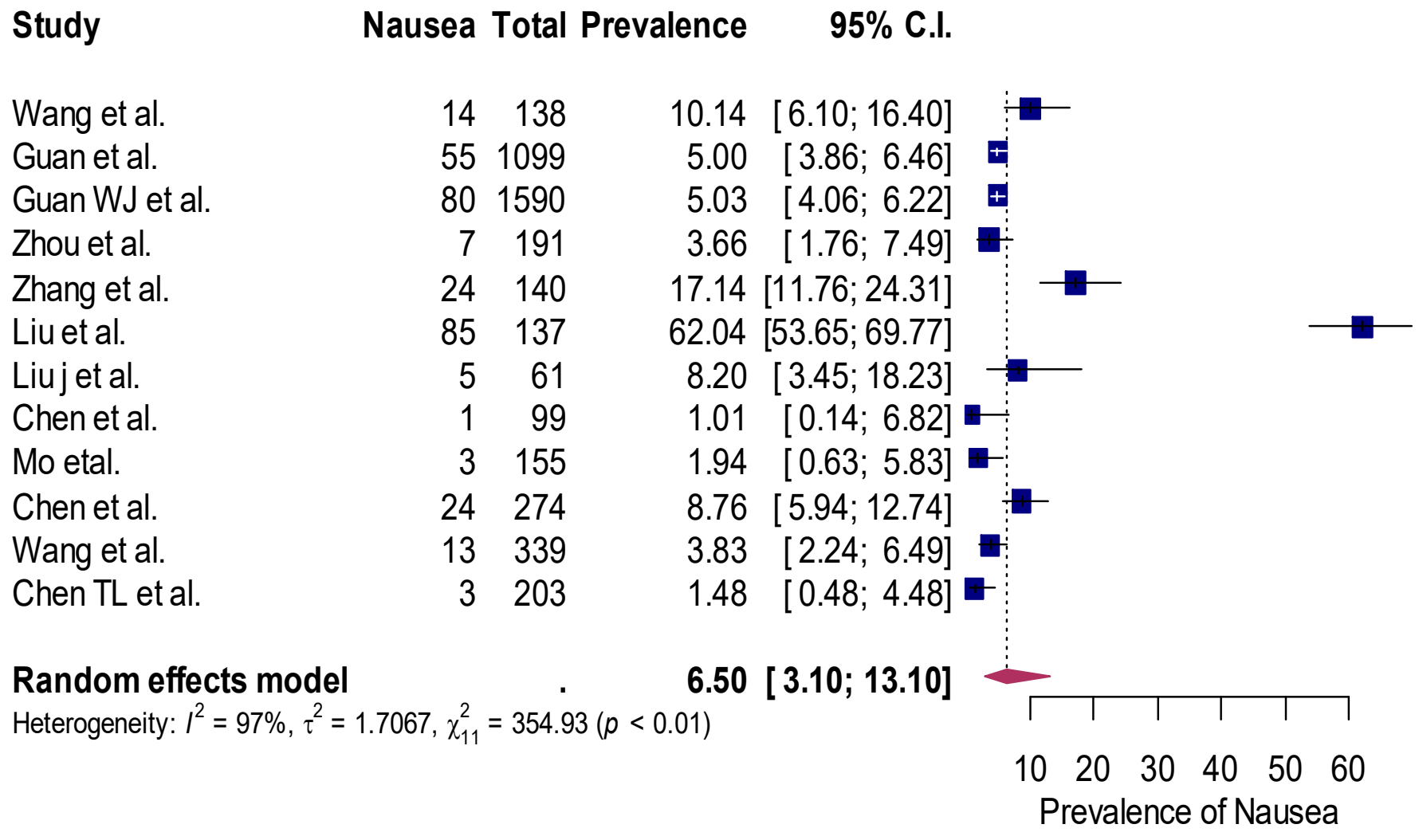

## I. Headache

| Study | Headache | Total | Prevalence | 95% C.I. |
|---|---|---|---|---|
| Wang et al. | 9 | 138 | 6.52 | [3.43; 12.06] |
| Xu et al. | 21 | 62 | 33.87 | [23.24; 46.43] |
| Guan et al. | 150 | 1099 | 13.65 | [11.74; 15.81] |
| Guan WJ et al. | 205 | 1590 | 12.89 | [11.33; 14.63] |
| Huang et al. | 3 | 41 | 7.32 | [2.38; 20.37] |
| Liu et al. | 13 | 137 | 9.49 | [5.59; 15.66] |
| Liu j et al. | 21 | 61 | 34.43 | [23.64; 47.10] |
| Chen et al. | 8 | 99 | 8.08 | [4.09; 15.33] |
| Yang et al. | 1 | 52 | 1.92 | [0.27; 12.43] |
| Mo etal. | 8 | 155 | 5.16 | [2.60; 9.98] |
| Du et al. | 4 | 85 | 4.71 | [1.78; 11.88] |
| Rong-Hui et al. | 17 | 179 | 9.50 | [5.99; 14.75] |
| Chen et al. | 31 | 274 | 11.31 | [8.07; 15.64] |
| Wang et al. | 12 | 339 | 3.54 | [2.02; 6.13] |
| Chen TL et al. | 10 | 203 | 4.93 | [2.67; 8.91] |
| **Random effects model** | | . | **9.84** | **[7.38; 13.00]** |

Heterogeneity: $I^2 = 87\%$, $\tau^2 = 0.2738$, $\chi^2_{14} = 103.80$ ($p < 0.01$)

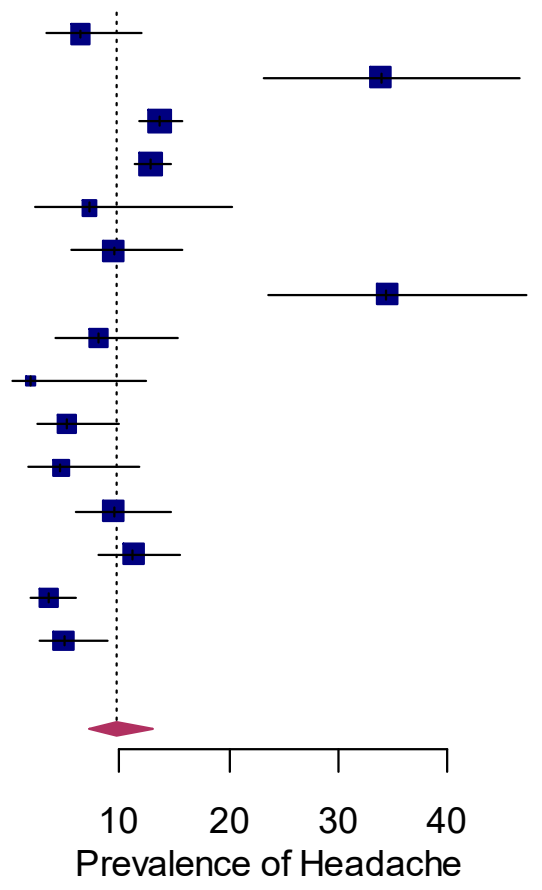

## J. Hypertension

| Study | Hypertension | Total | Prevalence | 95% C.I. |
|---|---|---|---|---|
| Wang et al. | 43 | 138 | 31.16 | [24.00; 39.35] |
| Richardson et al. | 3026 | 5700 | 53.09 | [51.79; 54.38] |
| Xu et al. | 5 | 62 | 8.06 | [3.40; 17.95] |
| Guan et al. | 165 | 1099 | 15.01 | [13.02; 17.25] |
| Guan WJ et al. | 269 | 1590 | 16.92 | [15.15; 18.84] |
| Huang et al. | 6 | 41 | 14.63 | [6.73; 28.96] |
| Guo et al. | 61 | 187 | 32.62 | [26.29; 39.66] |
| Zhou et al. | 58 | 191 | 30.37 | [24.26; 37.25] |
| Zhang et al. | 42 | 140 | 30.00 | [22.99; 38.09] |
| Liu et al. | 13 | 137 | 9.49 | [5.59; 15.66] |
| Liu j et al. | 12 | 61 | 19.67 | [11.53; 31.53] |
| Wu et al. | 39 | 201 | 19.40 | [14.51; 25.46] |
| Jie Li et al. | 1 | 17 | 5.88 | [0.82; 32.03] |
| Liu W et al. | 8 | 78 | 10.26 | [5.21; 19.19] |
| Mo etal. | 37 | 155 | 23.87 | [17.81; 31.21] |
| Du et al. | 32 | 85 | 37.65 | [28.02; 48.36] |
| Rong-Hui et al. | 58 | 179 | 32.40 | [25.95; 39.60] |
| Chen et al. | 93 | 274 | 33.94 | [28.58; 39.75] |
| Grasselli et al. | 509 | 1591 | 31.99 | [29.75; 34.33] |
| Deng et al. | 40 | 225 | 17.78 | [13.32; 23.33] |
| Wang et al. | 138 | 339 | 40.71 | [35.60; 46.02] |
| Chen TL et al. | 43 | 203 | 21.18 | [16.10; 27.34] |
| **Random effects model** | | . | **23.41** | **[17.53; 30.53]** |

Heterogeneity: $I^2 = 98\%$, $\tau^2 = 0.6866$, $\chi^2_{21} = 1252.84$ ($p < 0.01$)

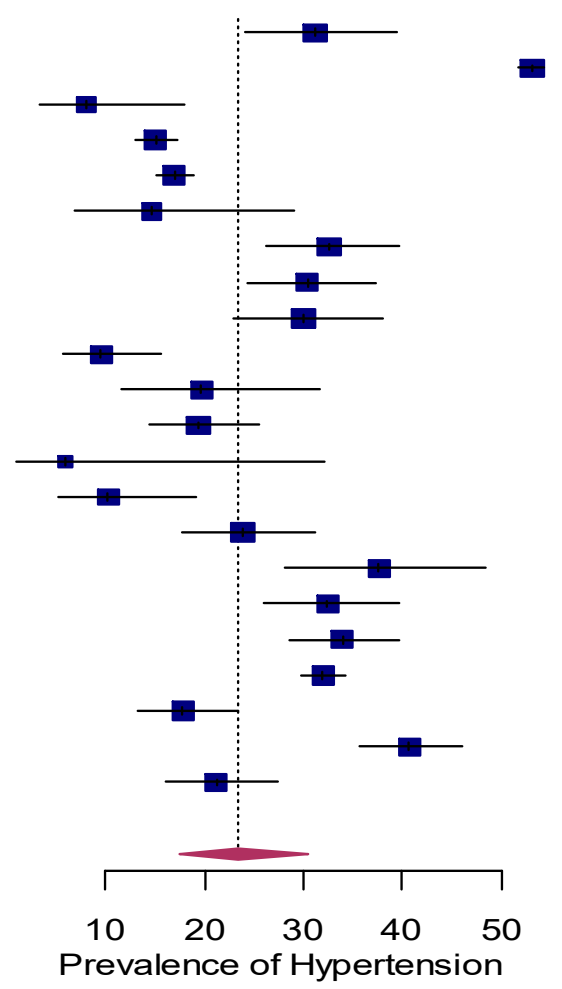

## K. Diabetes

| Study | Diabetes | Total | Prevalence | 95% C.I. |
|---|---|---|---|---|
| Wang et al. | 14 | 138 | 10.14 | [6.10; 16.40] |
| Richardson et al. | 1808 | 5700 | 31.72 | [30.52; 32.94] |
| Xu et al. | 1 | 62 | 1.61 | [0.23; 10.57] |
| Guan et al. | 81 | 1099 | 7.37 | [5.97; 9.07] |
| Guan WJ et al. | 130 | 1590 | 8.18 | [6.93; 9.63] |
| Huang et al. | 8 | 41 | 19.51 | [10.07; 34.42] |
| Guo et al. | 28 | 187 | 14.97 | [10.54; 20.83] |
| Zhou et al. | 36 | 191 | 18.85 | [13.91; 25.02] |
| Zhang et al. | 17 | 140 | 12.14 | [7.68; 18.67] |
| Liu et al. | 14 | 137 | 10.22 | [6.15; 16.52] |
| Liu j et al. | 5 | 61 | 8.20 | [3.45; 18.23] |
| Chen et al. | 12 | 99 | 12.12 | [7.01; 20.14] |
| Yang et al. | 2 | 52 | 3.85 | [0.96; 14.12] |
| Wu et al. | 22 | 201 | 10.95 | [7.32; 16.06] |
| Liu W et al. | 5 | 78 | 6.41 | [2.69; 14.49] |
| Mo etal. | 15 | 155 | 9.68 | [5.92; 15.43] |
| Du et al. | 19 | 85 | 22.35 | [14.74; 32.41] |
| Rong-Hui et al. | 33 | 179 | 18.44 | [13.41; 24.80] |
| Chen et al. | 47 | 274 | 17.15 | [13.14; 22.09] |
| Grasselli et al. | 180 | 1591 | 11.31 | [9.85; 12.97] |
| Deng et al. | 17 | 225 | 7.56 | [4.75; 11.82] |
| Wang et al. | 54 | 339 | 15.93 | [12.41; 20.22] |
| Chen TL et al. | 16 | 203 | 7.88 | [4.88; 12.48] |
| **Random effects model** | | . | **11.64** | **[8.27; 16.14]** |

Heterogeneity: $I^2 = 97\%$, $\tau^2 = 0.7625$, $\chi^2_{22} = 838.27$ ($p < 0.01$)

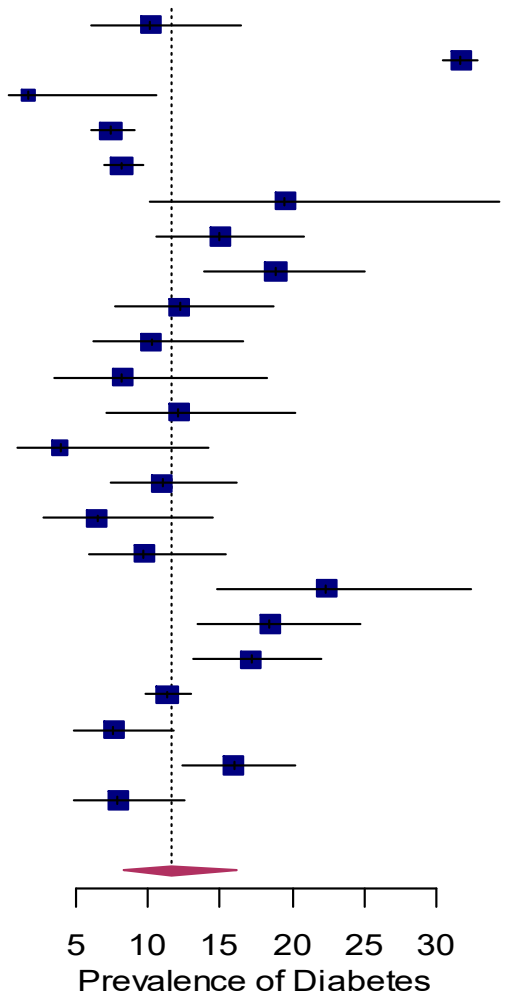

## L. CVD

| Study | CVD | Total | Prevalence | 95% C.I. |
|---|---|---|---|---|
| Wang et al. | 20 | 138 | 14.49 | [9.54; 21.40] |
| Richardson et al. | 824 | 5700 | 14.46 | [13.57; 15.39] |
| Guan et al. | 27 | 1099 | 2.46 | [1.69; 3.56] |
| Guan WJ et al. | 59 | 1590 | 3.71 | [2.89; 4.76] |
| Huang et al. | 2 | 41 | 4.88 | [1.22; 17.52] |
| Guo et al. | 21 | 187 | 11.23 | [7.44; 16.61] |
| Zhou et al. | 15 | 191 | 7.85 | [4.79; 12.62] |
| Zhang et al. | 10 | 140 | 7.14 | [3.89; 12.77] |
| Wu et al. | 25 | 80 | 31.25 | [22.08; 42.17] |
| Liu et al. | 10 | 137 | 7.30 | [3.97; 13.03] |
| Chen et al. | 40 | 99 | 40.40 | [31.21; 50.32] |
| Yang et al. | 4 | 52 | 7.69 | [2.92; 18.77] |
| Wu et al. | 8 | 201 | 3.98 | [2.00; 7.76] |
| Mo etal. | 15 | 155 | 9.68 | [5.92; 15.43] |
| Du et al. | 10 | 85 | 11.76 | [6.45; 20.50] |
| Rong-Hui et al. | 29 | 179 | 16.20 | [11.50; 22.34] |
| Chen et al. | 24 | 274 | 8.76 | [5.94; 12.74] |
| Grasselli et al. | 223 | 1591 | 14.02 | [12.40; 15.81] |
| Deng et al. | 13 | 225 | 5.78 | [3.38; 9.69] |
| Wang et al. | 53 | 339 | 15.63 | [12.14; 19.90] |
| Chen TL et al. | 16 | 203 | 7.88 | [4.88; 12.48] |
| **Random effects model** | | . | **10.00** | **[7.68; 12.93]** |

Heterogeneity: $I^2 = 94\%$, $\tau^2 = 0.3830$, $\chi^2_{20} = 328.46$ ($p < 0.01$)

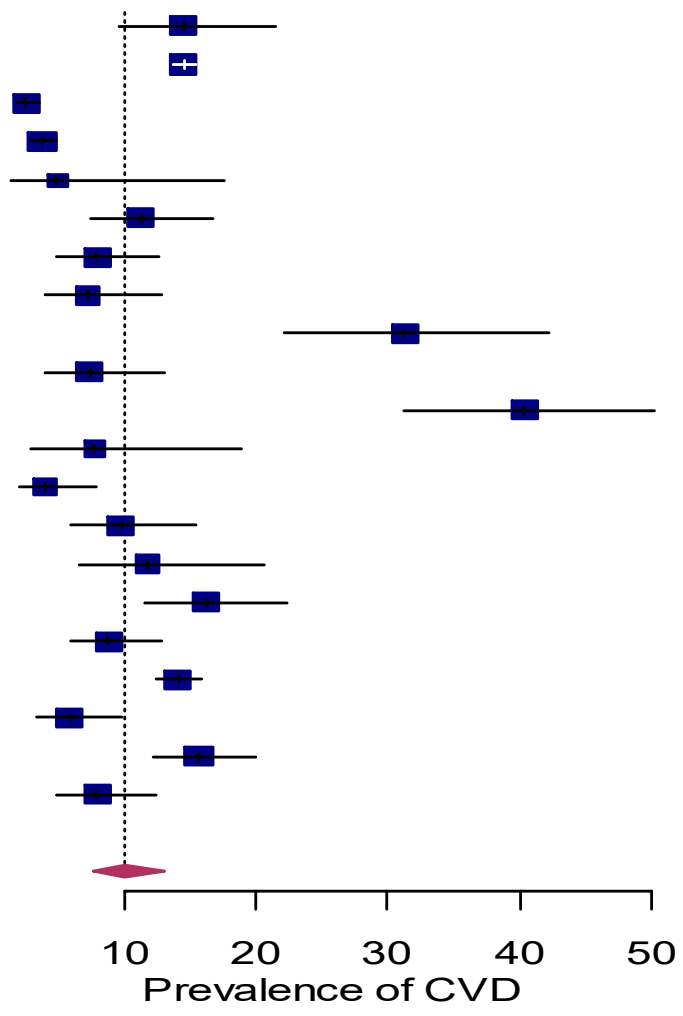

## M. Malignancy

| Study | Malignancy | Total | Prevalence | 95% C.I. |
|---|---|---|---|---|
| Wang et al. | 10 | 138 | 7.25 | [3.94; 12.94] |
| Richardson et al. | 320 | 5700 | 5.61 | [5.05; 6.24] |
| Guan et al. | 10 | 1099 | 0.91 | [0.49; 1.68] |
| Guan WJ et al. | 130 | 1590 | 8.18 | [6.93; 9.63] |
| Huang et al. | 1 | 41 | 2.44 | [0.34; 15.39] |
| Wu et al. | 4 | 80 | 5.00 | [1.89; 12.58] |
| Liu et al. | 2 | 137 | 1.46 | [0.37; 5.65] |
| Yang et al. | 1 | 52 | 1.92 | [0.27; 12.43] |
| Liu W et al. | 4 | 78 | 5.13 | [1.94; 12.88] |
| Mo etal. | 7 | 155 | 4.52 | [2.17; 9.17] |
| Du et al. | 6 | 85 | 7.06 | [3.21; 14.83] |
| Rong-Hui et al. | 4 | 179 | 2.23 | [0.84; 5.80] |
| Chen et al. | 7 | 274 | 2.55 | [1.22; 5.26] |
| Grasselli et al. | 81 | 1591 | 5.09 | [4.11; 6.29] |
| Deng et al. | 6 | 225 | 2.67 | [1.20; 5.81] |
| Wang et al. | 15 | 339 | 4.42 | [2.68; 7.21] |
| Chen TL et al. | 7 | 203 | 3.45 | [1.65; 7.05] |
| **Random effects model** | . | | **4.09** | **[3.18; 5.24]** |

Heterogeneity: $I^2 = 79\%$, $\tau^2 = 0.1566$, $\chi^2_{16} = 76.37$ ($p < 0.01$)

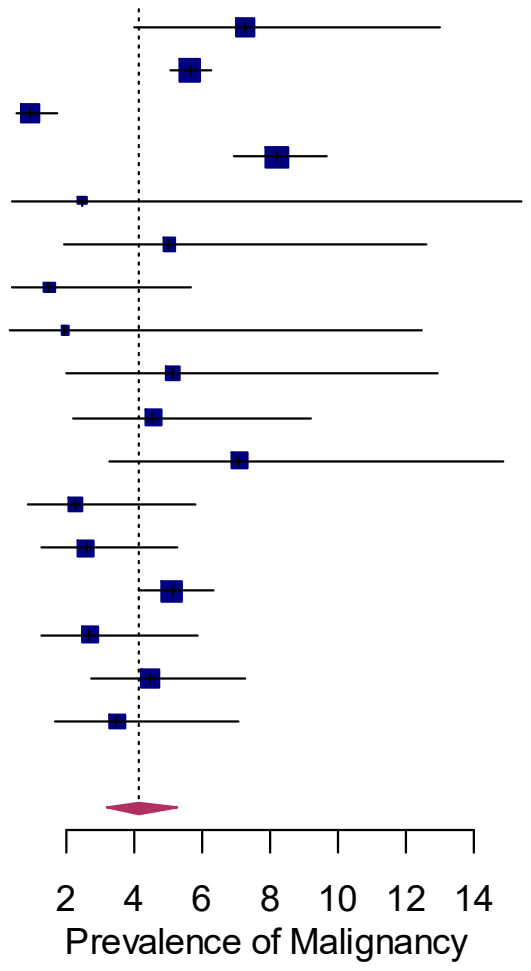

## N. COPD

| Study | COPD | Total | Prevalence | 95% C.I. |
|---|---|---|---|---|
| Wang et al. | 4 | 138 | 2.90 | [1.09; 7.47] |
| Richardson et al. | 287 | 5700 | 5.04 | [4.50; 5.63] |
| Xu et al. | 1 | 62 | 1.61 | [0.23; 10.57] |
| Guan et al. | 12 | 1099 | 1.09 | [0.62; 1.91] |
| Guan WJ et al. | 24 | 1590 | 1.51 | [1.01; 2.24] |
| Huang et al. | 1 | 41 | 2.44 | [0.34; 15.39] |
| Guo et al. | 4 | 187 | 2.14 | [0.81; 5.56] |
| Zhang et al. | 2 | 140 | 1.43 | [0.36; 5.53] |
| Liu et al. | 2 | 137 | 1.46 | [0.37; 5.65] |
| Liu j et al. | 5 | 61 | 8.20 | [3.45; 18.23] |
| Chen et al. | 1 | 99 | 1.01 | [0.14; 6.82] |
| Wu et al. | 5 | 201 | 2.49 | [1.04; 5.84] |
| Liu W et al. | 2 | 78 | 2.56 | [0.64; 9.68] |
| Mo etal. | 5 | 155 | 3.23 | [1.35; 7.51] |
| Du et al. | 2 | 85 | 2.35 | [0.59; 8.92] |
| Chen et al. | 18 | 274 | 6.57 | [4.18; 10.18] |
| Grasselli et al. | 42 | 1591 | 2.64 | [1.96; 3.55] |
| Deng et al. | 22 | 225 | 9.78 | [6.52; 14.40] |
| Wang et al. | 21 | 339 | 6.19 | [4.07; 9.31] |
| Chen TL et al. | 8 | 203 | 3.94 | [1.98; 7.68] |
| **Random effects model** | . | | **3.18** | **[2.33; 4.34]** |

Heterogeneity: $I^2 = 82\%$, $\tau^2 = 0.3245$, $\chi^2_{19} = 106.99$ ($p < 0.01$)

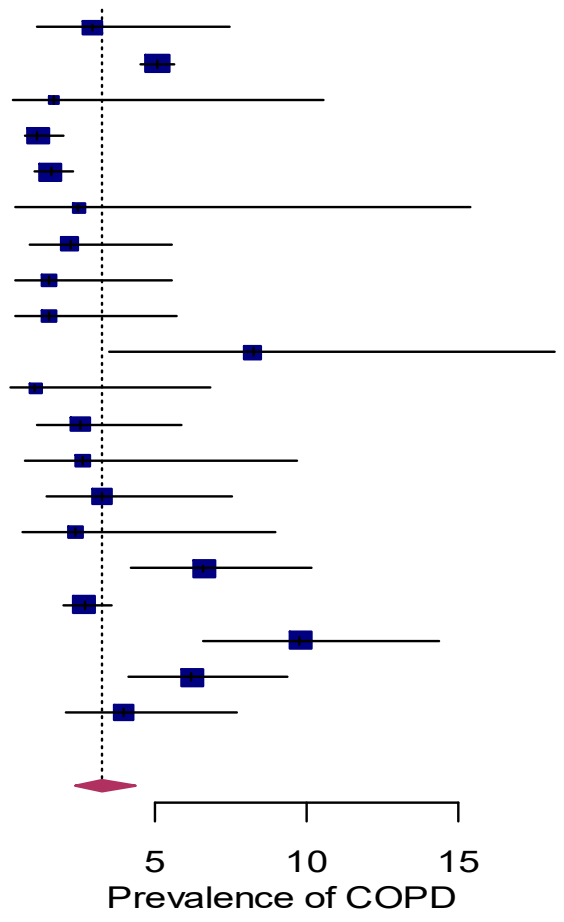

## O. CERD

| Study | CERD | Total | Prevalence | 95% C.I. |
|---|---|---|---|---|
| Wang et al. | 7 | 138 | 5.07 | [2.44; 10.26] |
| Xu et al. | 1 | 62 | 1.61 | [0.23; 10.57] |
| Guan et al. | 15 | 1099 | 1.36 | [0.82; 2.25] |
| Guan WJ et al. | 30 | 1590 | 1.89 | [1.32; 2.69] |
| Liu j et al. | 1 | 61 | 1.64 | [0.23; 10.73] |
| Mo etal. | 7 | 155 | 4.52 | [2.17; 9.17] |
| Du et al. | 7 | 85 | 8.24 | [3.98; 16.28] |
| Chen et al. | 4 | 274 | 1.46 | [0.55; 3.82] |
| Wang et al. | 21 | 339 | 6.19 | [4.07; 9.31] |
| Chen TL et al. | 9 | 203 | 4.43 | [2.32; 8.30] |
| **Random effects model** | . | | **3.23** | **[2.02; 5.13]** |

Heterogeneity: $I^2 = 79\%$, $\tau^2 = 0.4118$, $\chi^2_{9} = 42.32$ ($p < 0.01$)

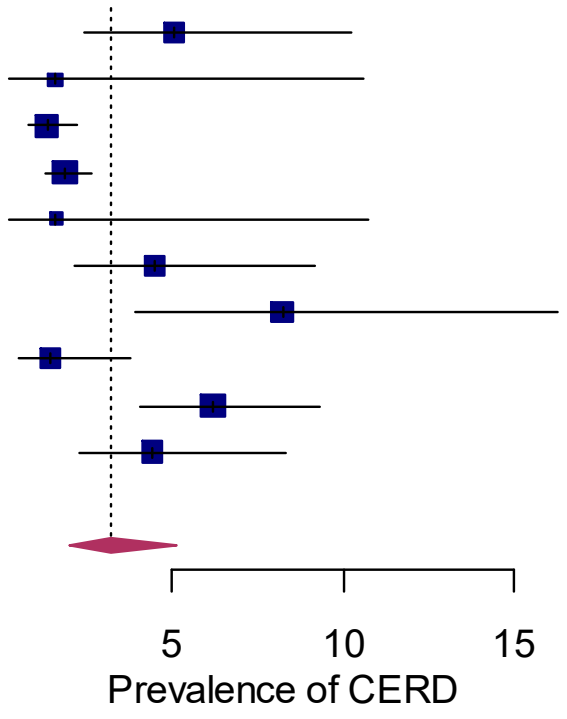

## P. CKD

| Study | CKD | Total | Prevalence | 95% C.I. |
|---|---|---|---|---|
| Wang et al. | 4 | 138 | 2.90 | [ 1.09; 7.47] |
| Richardson et al. | 453 | 5700 | 7.95 | [ 7.27; 8.68] |
| Xu et al. | 1 | 62 | 1.61 | [ 0.23; 10.57] |
| Guan et al. | 8 | 1099 | 0.73 | [ 0.36; 1.45] |
| Guan WJ et al. | 269 | 1590 | 16.92 | [15.15; 18.84] |
| Guo et al. | 6 | 187 | 3.21 | [ 1.45; 6.96] |
| Zhou et al. | 2 | 191 | 1.05 | [ 0.26; 4.09] |
| Zhang et al. | 2 | 140 | 1.43 | [ 0.36; 5.53] |
| Wu et al. | 1 | 80 | 1.25 | [ 0.18; 8.34] |
| Wu et al. | 2 | 201 | 1.00 | [ 0.25; 3.89] |
| Mo etal. | 6 | 155 | 3.87 | [ 1.75; 8.35] |
| Du et al. | 3 | 85 | 3.53 | [ 1.14; 10.38] |
| Rong-Hui et al. | 4 | 179 | 2.23 | [ 0.84; 5.80] |
| Chen et al. | 4 | 274 | 1.46 | [ 0.55; 3.82] |
| Grasselli et al. | 36 | 1591 | 2.26 | [ 1.64; 3.12] |
| Wang et al. | 13 | 339 | 3.83 | [ 2.24; 6.49] |
| Chen TL et al. | 8 | 203 | 3.94 | [ 1.98; 7.68] |
| **Random effects model** | . | | **2.78** | **[ 1.74; 4.41]** |

Heterogeneity: $I^2 = 95\%$, $\tau^2 = 0.7596$, $\chi^2_{16} = 344.41$ ($p < 0.01$)

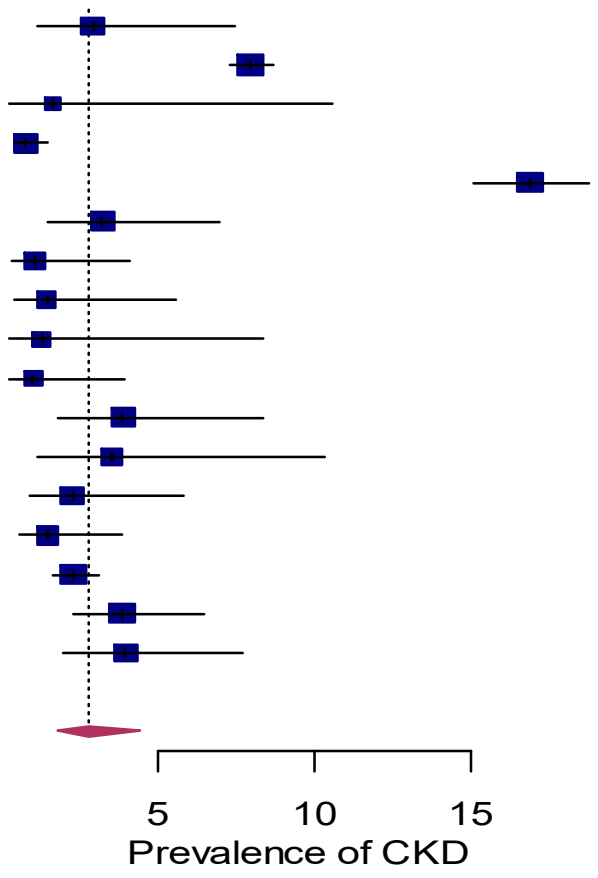

## Q. CLD

| Study | CLD | Total | Prevalence | 95% C.I. |
|---|---|---|---|---|
| Wang et al. | 4 | 138 | 2.90 | [1.09; 7.47] |
| Richardson et al. | 11 | 5700 | 0.19 | [0.11; 0.35] |
| Xu et al. | 7 | 62 | 11.29 | [5.48; 21.84] |
| Guan et al. | 23 | 1099 | 2.09 | [1.39; 3.13] |
| Guan WJ et al. | 24 | 1590 | 1.51 | [1.01; 2.24] |
| Huang et al. | 1 | 41 | 2.44 | [0.34; 15.39] |
| Zhang et al. | 8 | 140 | 5.71 | [2.88; 11.01] |
| Wu et al. | 1 | 80 | 1.25 | [0.18; 8.34] |
| Wu et al. | 7 | 201 | 3.48 | [1.67; 7.12] |
| Mo etal. | 7 | 155 | 4.52 | [2.17; 9.17] |
| Du et al. | 5 | 85 | 5.88 | [2.47; 13.36] |
| Chen et al. | 11 | 274 | 4.01 | [2.24; 7.10] |
| Grasselli et al. | 28 | 1591 | 1.76 | [1.22; 2.54] |
| Wang et al. | 2 | 339 | 0.59 | [0.15; 2.33] |
| Chen TL et al. | 8 | 203 | 3.94 | [1.98; 7.68] |
| **Random effects model** | | . | **2.50** | **[1.51; 4.11]** |

Heterogeneity: $I^2 = 88\%$, $\tau^2 = 0.8257$, $\chi^2_{14} = 118.20$ ($p$ < 0.01)

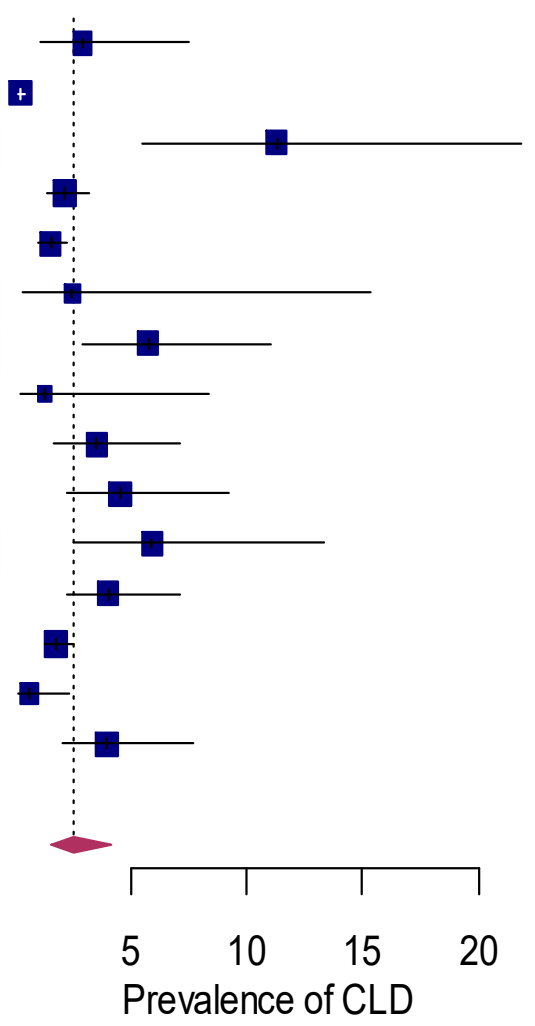

## R. Smoking

| Study | Smoking | Total | Prevalence | 95% C.I. |
|---|---|---|---|---|
| Richardson et al. | 2691 | 5700 | 47.21 | [45.92; 48.51] |
| Guan et al. | 158 | 1099 | 14.38 | [12.42; 16.58] |
| Guan WJ et al. | 111 | 1590 | 6.98 | [ 5.83; 8.34] |
| Huang et al. | 3 | 41 | 7.32 | [ 2.38; 20.37] |
| Guo et al. | 18 | 187 | 9.63 | [ 6.15; 14.76] |
| Zhou et al. | 11 | 191 | 5.76 | [ 3.22; 10.10] |
| Zhang et al. | 9 | 140 | 6.43 | [ 3.38; 11.89] |
| Liu j et al. | 4 | 61 | 6.56 | [ 2.48; 16.21] |
| Yang et al. | 2 | 52 | 3.85 | [ 0.96; 14.12] |
| Jie Li et al. | 3 | 17 | 17.65 | [ 5.80; 42.71] |
| Liu W et al. | 5 | 78 | 6.41 | [ 2.69; 14.49] |
| Mo etal. | 6 | 155 | 3.87 | [ 1.75; 8.35] |
| Feng et al. | 44 | 476 | 9.24 | [ 6.95; 12.20] |
| Chen et al. | 19 | 274 | 6.93 | [ 4.47; 10.61] |
| **Random effects model** | | . | **8.83** | **[ 4.19; 17.69]** |

Heterogeneity: $I^2 = 99\%$, $\tau^2 = 2.1578$, $\chi^2_{13} = 1345.04$ ($p$ < 0.01)

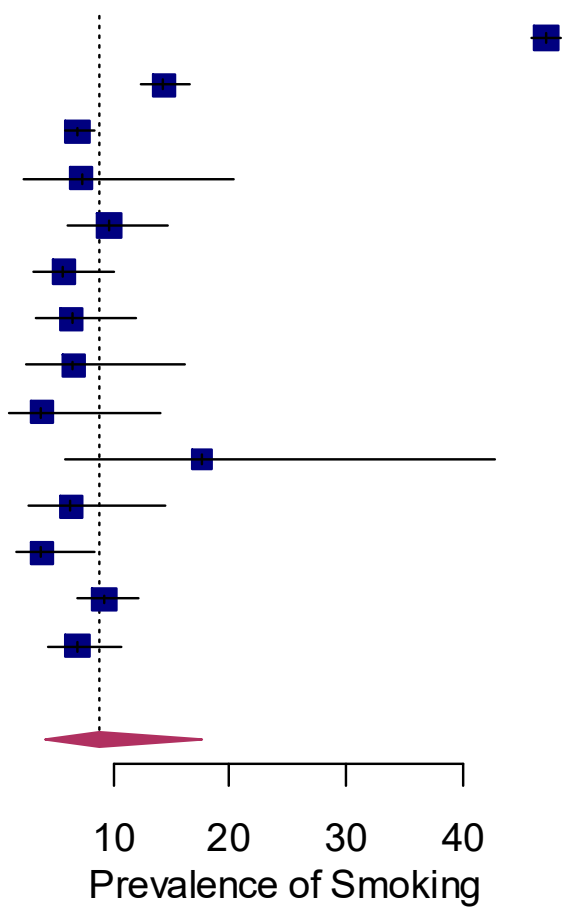

**Supplementary Figure 1: Meta-analysis of prevalence of comorbidities and symptoms COVID-19 fatalities.**

## A. Fever

| Study | Experimental Events | Experimental Total | Control Events | Control Total | Weight | Odds Ratio MH, Random, 95% CI |
|---|---|---|---|---|---|---|
| Wang et al. | 36 | 136 | 36 | 138 | 9.1% | 1.02 [0.60; 1.75] |
| Guan et al. | 82 | 473 | 173 | 1099 | 16.3% | 1.12 [0.84; 1.50] |
| Huang et al. | 13 | 40 | 13 | 41 | 4.0% | 1.04 [0.41; 2.64] |
| Zhou et al. | 51 | 180 | 66 | 191 | 11.4% | 0.75 [0.48; 1.16] |
| Zhang et al. | 51 | 110 | 58 | 140 | 9.8% | 1.22 [0.74; 2.02] |
| Liu et al. | 17 | 60 | 17 | 61 | 5.2% | 1.02 [0.46; 2.26] |
| Yang et al. | 31 | 51 | 52 | 52 | 0.5% | 0.01 [0.00; 0.25] |
| Mo etal. | 63 | 126 | 55 | 155 | 10.4% | 1.82 [1.13; 2.94] |
| Chen et al. | 104 | 249 | 113 | 274 | 14.2% | 1.02 [0.72; 1.45] |
| Wang et al. | 56 | 311 | 65 | 339 | 12.7% | 0.93 [0.62; 1.38] |
| Chen TL et al. | 17 | 181 | 19 | 203 | 6.5% | 1.00 [0.50; 2.00] |
| **Total (95% CI)** | | **1917** | | **2693** | **100.0%** | **1.04 [0.85; 1.28]** |
| **Prediction interval** | | | | | | **[0.61; 1.79]** |

Heterogeneity: $Tau^2 = 0.0459$; $Chi^2 = 17.34$, df = 10 (P = 0.07); $I^2 = 42\%$

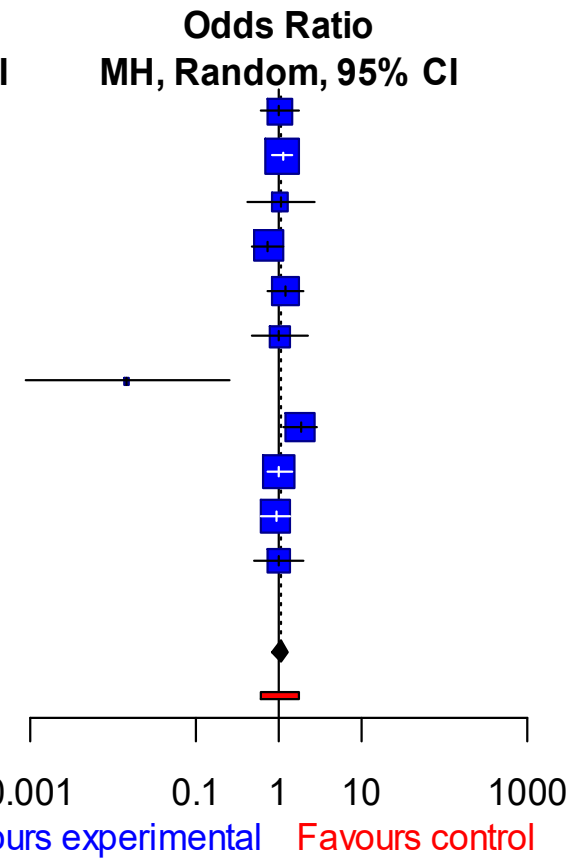

## B. Cough

| Study | Experimental Events | Experimental Total | Control Events | Control Total | Weight | Odds Ratio MH, Random, 95% CI |
|---|---|---|---|---|---|---|
| Wang et al. | 21 | 82 | 36 | 138 | 8.0% | 0.98 [0.52; 1.82] |
| Guan et al. | 122 | 745 | 173 | 1099 | 19.4% | 1.05 [0.81; 1.35] |
| Huang et al. | 11 | 31 | 13 | 41 | 3.9% | 1.18 [0.44; 3.18] |
| Zhou et al. | 39 | 151 | 66 | 191 | 11.5% | 0.66 [0.41; 1.06] |
| Zhang et al. | 45 | 90 | 58 | 140 | 9.9% | 1.41 [0.83; 2.41] |
| Liu et al. | 14 | 39 | 17 | 61 | 4.9% | 1.45 [0.61; 3.43] |
| Yang et al. | 25 | 15 | 52 | 52 | 0.0% | |
| Mo etal. | 54 | 97 | 55 | 155 | 10.2% | 2.28 [1.36; 3.83] |
| Chen et al. | 79 | 185 | 113 | 274 | 14.4% | 1.06 [0.73; 1.55] |
| Wang et al. | 30 | 179 | 65 | 339 | 11.3% | 0.85 [0.53; 1.37] |
| Chen TL et al. | 15 | 122 | 19 | 203 | 6.5% | 1.36 [0.66; 2.78] |
| **Total (95% CI)** | | **1736** | | **2693** | **100.0%** | **1.12 [0.91; 1.38]** |
| **Prediction interval** | | | | | | **[0.65; 1.92]** |

Heterogeneity: $Tau^2 = 0.0430$; $Chi^2 = 15.20$, df = 9 (P = 0.09); $I^2 = 41\%$

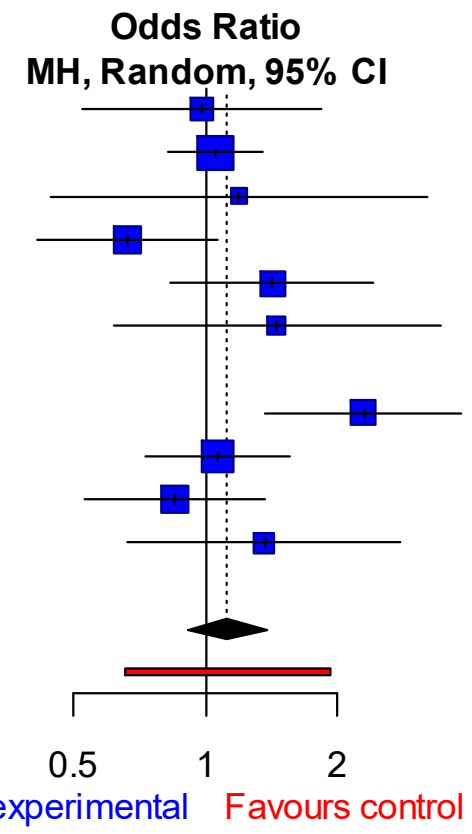

## C. Fatigue

| Study | Experimental Events | Experimental Total | Control Events | Control Total | Weight | Odds Ratio MH, Random, 95% CI |
|---|---|---|---|---|---|---|
| Wang et al. | 29 | 96 | 36 | 138 | 8.7% | 1.23 [0.69; 2.19] |
| Guan et al. | 69 | 419 | 173 | 1099 | 31.4% | 1.06 [0.78; 1.43] |
| Huang et al. | 7 | 18 | 13 | 41 | 2.2% | 1.37 [0.43; 4.34] |
| Zhou et al. | 15 | 44 | 66 | 191 | 6.1% | 0.98 [0.49; 1.96] |
| Zhang et al. | 39 | 90 | 58 | 140 | 10.2% | 1.08 [0.63; 1.85] |
| Liu j et al. | 12 | 35 | 17 | 61 | 3.6% | 1.35 [0.55; 3.30] |
| Mo etal. | 27 | 60 | 55 | 155 | 8.0% | 1.49 [0.81; 2.73] |
| Chen et al. | 64 | 137 | 113 | 274 | 17.1% | 1.25 [0.83; 1.89] |
| Wang et al. | 26 | 135 | 65 | 339 | 11.4% | 1.01 [0.61; 1.67] |
| Chen TL et al. | 2 | 16 | 19 | 203 | 1.2% | 1.38 [0.29; 6.55] |
| **Total (95% CI)** | | **1050** | | **2641** | **100.0%** | **1.14 [0.96; 1.36]** |
| **Prediction interval** | | | | | | **[0.93; 1.40]** |

Heterogeneity: $Tau^2 = 0$; $Chi^2 = 1.99$, df = 9 (P = 0.99); $I^2 = 0\%$

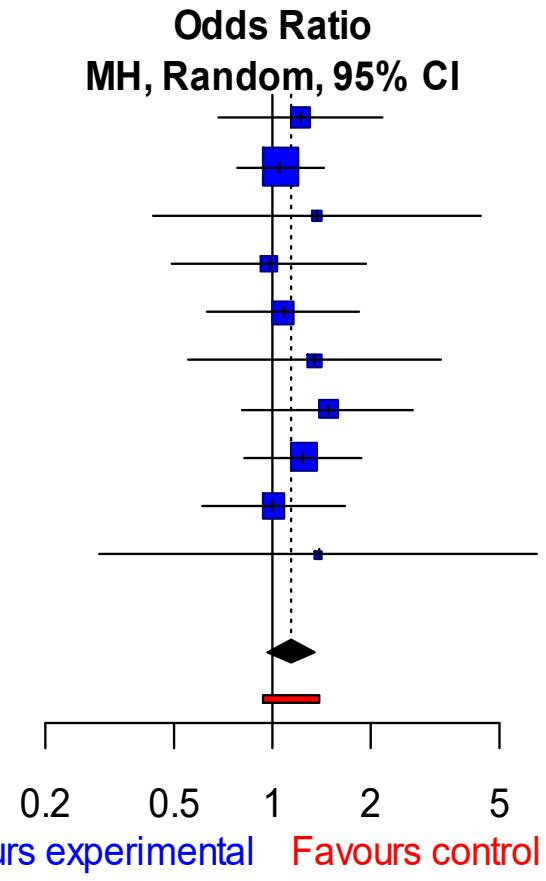

## D. Anorexia

| Study | Experimental Events | Experimental Total | Control Events | Control Total | Weight | Odds Ratio MH, Random, 95% CI |
|---|---|---|---|---|---|---|
| Wang et al. | 24 | 55 | 36 | 138 | 21.7% | 2.19 [1.14; 4.22] |
| Zhang et al. | 8 | 17 | 58 | 140 | 14.7% | 1.26 [0.46; 3.45] |
| Mo etal. | 18 | 26 | 55 | 155 | 16.7% | 4.09 [1.67; 10.02] |
| Chen et al. | 31 | 66 | 113 | 274 | 24.4% | 1.26 [0.74; 2.17] |
| Wang et al. | 15 | 94 | 65 | 339 | 22.6% | 0.80 [0.43; 1.48] |
| **Total (95% CI)** | | **258** | | **1046** | **100.0%** | **1.56 [0.93; 2.62]** |
| **Prediction interval** | | | | | | **[0.29; 8.45]** |

Heterogeneity: $Tau^2 = 0.2118$; $Chi^2 = 10.61$, df = 4 (P = 0.03); $I^2 = 62\%$

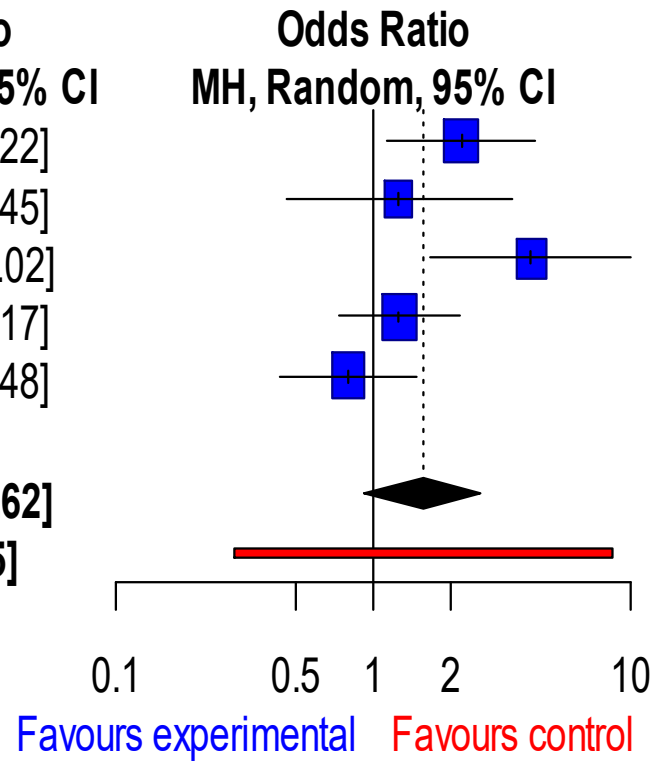

## E. Myalgia

| Study | Experimental Events | Experimental Total | Control Events | Control Total | Weight | Odds Ratio MH, Random, 95% CI |
|---|---|---|---|---|---|---|
| Wang et al. | 12 | 48 | 36 | 138 | 23.4% | 0.94 [0.44; 2.01] |
| Huang et al. | 7 | 18 | 13 | 41 | 10.1% | 1.37 [0.43; 4.34] |
| Zhou et al. | 8 | 29 | 66 | 191 | 17.8% | 0.72 [0.30; 1.72] |
| Yang et al. | 4 | 2 | 52 | 52 | 0.0% | |
| Chen et al. | 21 | 60 | 113 | 274 | 39.5% | 0.77 [0.43; 1.37] |
| Wang et al. | 1 | 16 | 65 | 339 | 3.2% | 0.28 [0.04; 2.17] |
| Chen TL et al. | 2 | 54 | 19 | 203 | 6.0% | 0.37 [0.08; 1.65] |
| **Total (95% CI)** | | **227** | | **1238** | **100.0%** | **0.78 [0.54; 1.13]** |
| **Prediction interval** | | | | | | **[0.47; 1.31]** |

Heterogeneity: $Tau^2 = 0$; $Chi^2 = 3.14$, df = 5 (P = 0.68); $I^2 = 0\%$

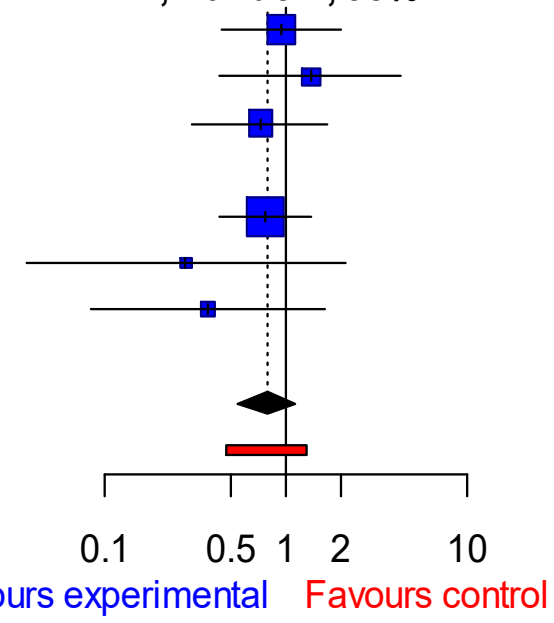

## F. Dyspnea

| Study | Experimental Events | Experimental Total | Control Events | Control Total | Weight | Odds Ratio MH, Random, 95% CI |
|---|---|---|---|---|---|---|
| Wang et al. | 23 | 43 | 36 | 138 | 29.0% | 3.26 [1.60; 6.62] |
| Huang et al. | 12 | 22 | 13 | 41 | 15.8% | 2.58 [0.89; 7.51] |
| Liu j et al. | 3 | 3 | 17 | 61 | 2.4% | 17.80 [0.87; 362.69] |
| Yang et al. | 21 | 12 | 52 | 52 | 0.0% | |
| Mo etal. | 2 | 2 | 55 | 155 | 2.3% | 9.05 [0.43; 191.95] |
| Wang et al. | 38 | 138 | 65 | 339 | 47.0% | 1.60 [1.01; 2.54] |
| Chen TL et al. | 1 | 3 | 19 | 203 | 3.5% | 4.84 [0.42; 55.91] |
| **Total (95% CI)** | | **223** | | **989** | **100.0%** | **2.43 [1.52; 3.89]** |
| **Prediction interval** | | | | | | **[0.92; 6.45]** |

Heterogeneity: $Tau^2 = 0.0664$; $Chi^2 = 6.17$, df = 5 (P = 0.29); $I^2 = 19\%$

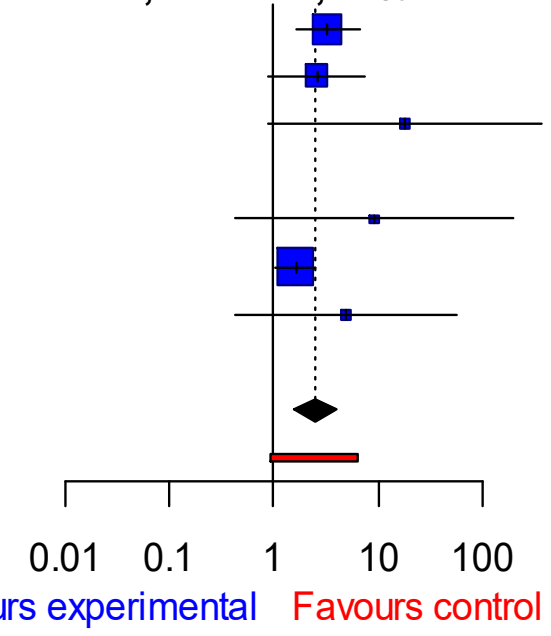

## G. Diarrhea

| Study | Experimental Events | Experimental Total | Control Events | Control Total | Weight | Odds Ratio MH, Random, 95% CI |
|---|---|---|---|---|---|---|
| Wang et al. | 6 | 14 | 36 | 138 | 8.7% | 2.12 [0.69; 6.54] |
| Guan et al. | 10 | 42 | 173 | 1099 | 19.0% | 1.67 [0.81; 3.47] |
| Zhou et al. | 2 | 9 | 66 | 191 | 4.4% | 0.54 [0.11; 2.68] |
| Zhang et al. | 9 | 18 | 58 | 140 | 11.1% | 1.41 [0.53; 3.78] |
| Liu j et al. | 1 | 6 | 17 | 61 | 2.3% | 0.52 [0.06; 4.76] |
| Mo etal. | 5 | 7 | 55 | 155 | 4.1% | 4.55 [0.85; 24.21] |
| Chen et al. | 27 | 77 | 113 | 274 | 32.2% | 0.77 [0.45; 1.30] |
| Wang et al. | 8 | 43 | 65 | 339 | 15.6% | 0.96 [0.43; 2.18] |
| Chen TL et al. | 1 | 10 | 19 | 203 | 2.6% | 1.08 [0.13; 8.96] |
| **Total (95% CI)** | | **226** | | **2600** | **100.0%** | **1.14 [0.81; 1.61]** |
| **Prediction interval** | | | | | | **[0.66; 1.96]** |

Heterogeneity: $Tau^2 = 0.0224$; $Chi^2 = 8.68$, df = 8 (P = 0.37); $I^2 = 8\%$

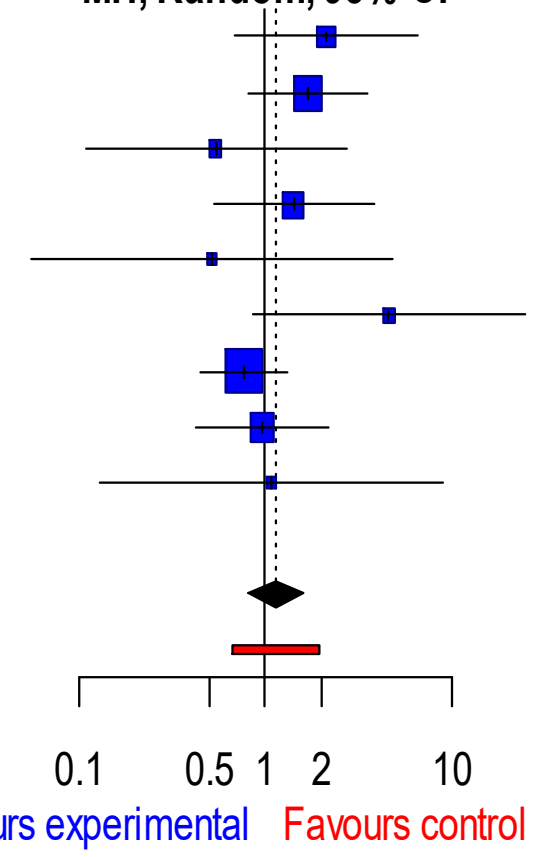

## H. Nausea

| Study | Experimental Events | Experimental Total | Control Events | Control Total | Weight | Odds Ratio MH, Random, 95% CI |
|---|---|---|---|---|---|---|
| Wang et al. | 4 | 14 | 36 | 138 | 12.3% | 1.13 [0.33; 3.84] |
| Guan et al. | 12 | 55 | 173 | 1099 | 31.9% | 1.49 [0.77; 2.89] |
| Zhou et al. | 3 | 7 | 66 | 191 | 8.3% | 1.42 [0.31; 6.54] |
| Zhang et al. | 5 | 24 | 58 | 140 | 16.1% | 0.37 [0.13; 1.05] |
| Liu j et al. | 2 | 5 | 17 | 61 | 5.6% | 1.73 [0.26; 11.25] |
| Chen et al. | 8 | 24 | 113 | 274 | 21.0% | 0.71 [0.29; 1.72] |
| Wang et al. | 1 | 13 | 65 | 339 | 4.7% | 0.35 [0.04; 2.75] |
| **Total (95% CI)** | | **142** | | **2242** | **100.0%** | **0.93 [0.58; 1.47]** |
| **Prediction interval** | | | | | | **[0.39; 2.20]** |

Heterogeneity: $Tau^2 = 0.0589$; $Chi^2 = 7.07$, df = 6 (P = 0.31); $I^2 = 15\%$

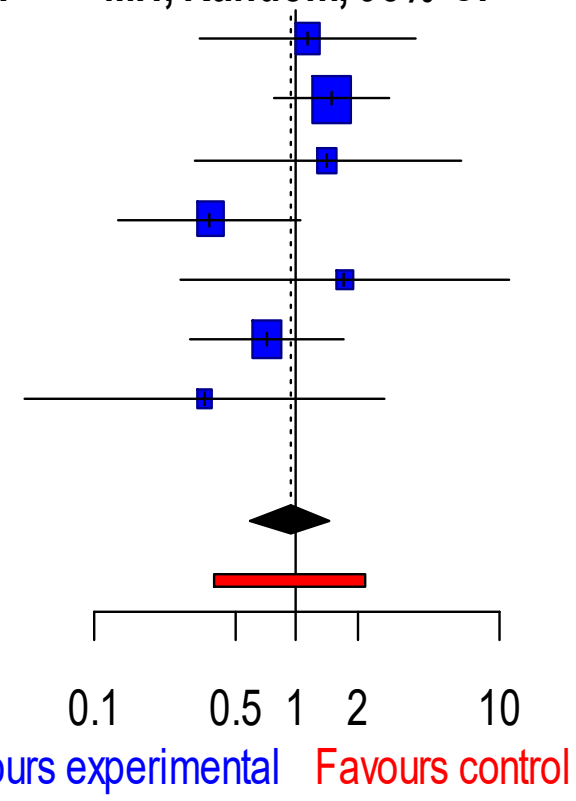

## I. Headache

| Study | Experimental Events | Experimental Total | Control Events | Control Total | Weight | Odds Ratio MH, Random, 95% CI |
|---|---|---|---|---|---|---|
| Wang et al. | 3 | 9 | 36 | 138 | 7.6% | 1.42 [0.34; 5.96] |
| Guan et al. | 26 | 150 | 173 | 1099 | 53.1% | 1.12 [0.71; 1.77] |
| Liu j et al. | 3 | 21 | 17 | 61 | 8.6% | 0.43 [0.11; 1.65] |
| Yang et al. | 2 | 1 | 52 | 52 | 0.0% | |
| Mo etal. | 5 | 8 | 55 | 155 | 7.3% | 3.03 [0.70; 13.16] |
| Chen et al. | 11 | 31 | 113 | 274 | 23.4% | 0.78 [0.36; 1.70] |
| **Total (95% CI)** | | **220** | | **1779** | **100.0%** | **1.04 [0.69; 1.56]** |
| **Prediction interval** | | | | | | **[0.45; 2.41]** |

Heterogeneity: $Tau^2 = 0.0272$; $Chi^2 = 4.49$, df = 4 (P = 0.34); $I^2 = 11\%$

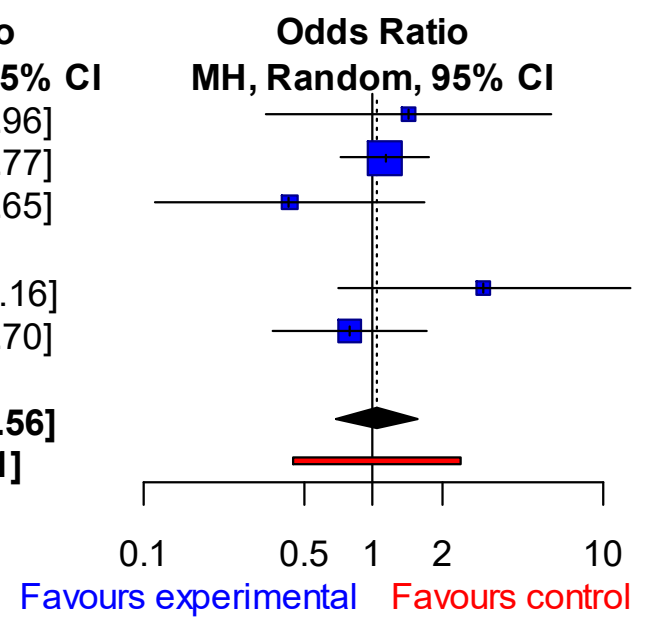

## J. Hypertension

| Study | Experimental Events | Experimental Total | Control Events | Control Total | Weight | Odds Ratio MH, Random, 95% CI |
|---|---|---|---|---|---|---|
| Wang et al. | 21 | 43 | 36 | 138 | 7.6% | 2.70 [1.33; 5.49] |
| Guan et al. | 41 | 165 | 173 | 1099 | 25.1% | 1.77 [1.20; 2.61] |
| Huang et al. | 2 | 6 | 13 | 41 | 1.1% | 1.08 [0.17; 6.65] |
| Zhou et al. | 26 | 58 | 66 | 191 | 10.6% | 1.54 [0.85; 2.80] |
| Zhang et al. | 22 | 42 | 58 | 140 | 7.9% | 1.56 [0.78; 3.11] |
| Liu j et al. | 6 | 12 | 17 | 61 | 2.4% | 2.59 [0.73; 9.15] |
| Mo etal. | 22 | 37 | 55 | 155 | 7.0% | 2.67 [1.28; 5.56] |
| Chen et al. | 54 | 93 | 113 | 274 | 16.7% | 1.97 [1.22; 3.18] |
| Wang et al. | 32 | 138 | 65 | 339 | 16.5% | 1.27 [0.79; 2.05] |
| Chen TL et al. | 9 | 43 | 19 | 203 | 5.0% | 2.56 [1.07; 6.14] |
| **Total (95% CI)** | | **637** | | **2641** | **100.0%** | **1.81 [1.49; 2.20]** |
| **Prediction interval** | | | | | | **[1.44; 2.27]** |

Heterogeneity: $Tau^2 = 0$; $Chi^2 = 6.22$, df = 9 (P = 0.72); $I^2 = 0\%$

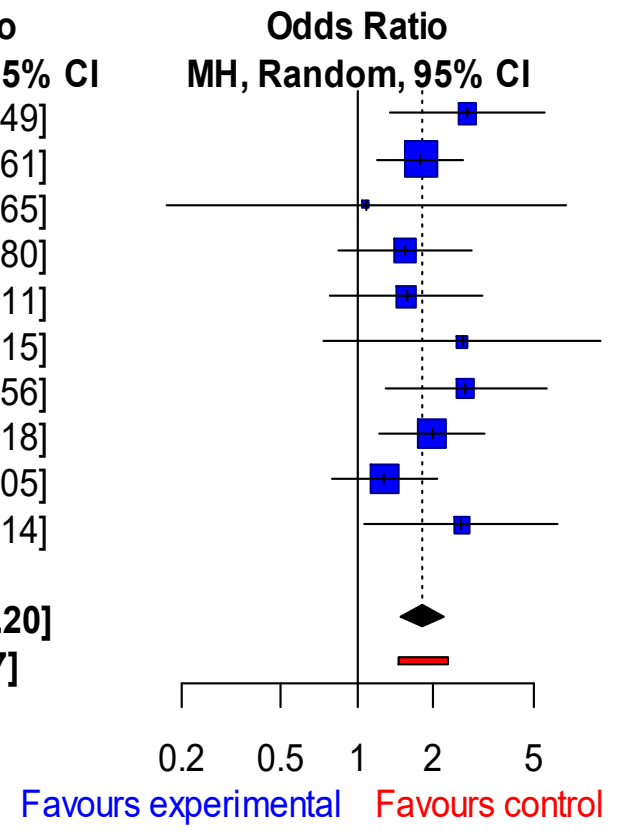

## K. Diabetes

| Study | Experimental Events | Experimental Total | Control Events | Control Total | Weight | Odds Ratio MH, Random, 95% CI |
|---|---|---|---|---|---|---|
| Wang et al. | 8 | 14 | 36 | 138 | 8.3% | 3.78 [1.23; 11.63] |
| Guan et al. | 28 | 81 | 173 | 1099 | 17.9% | 2.83 [1.74; 4.60] |
| Huang et al. | 1 | 8 | 13 | 41 | 2.9% | 0.31 [0.03; 2.77] |
| Zhou et al. | 17 | 36 | 66 | 191 | 13.6% | 1.69 [0.83; 3.48] |
| Zhang et al. | 8 | 17 | 58 | 140 | 9.5% | 1.26 [0.46; 3.45] |
| Liu j et al. | 3 | 5 | 17 | 61 | 3.8% | 3.88 [0.60; 25.31] |
| Yang et al. | 7 | 2 | 52 | 52 | 0.0% | |
| Mo etal. | 12 | 15 | 55 | 155 | 6.8% | 7.27 [1.97; 26.88] |
| Chen et al. | 24 | 47 | 113 | 274 | 15.4% | 1.49 [0.80; 2.76] |
| Wang et al. | 11 | 54 | 65 | 339 | 13.7% | 1.08 [0.53; 2.21] |
| Chen TL et al. | 5 | 16 | 19 | 203 | 8.0% | 4.40 [1.38; 14.01] |
| **Total (95% CI)** | | **295** | | **2693** | **100.0%** | **2.08 [1.39; 3.10]** |
| **Prediction interval** | | | | | | **[0.72; 6.01]** |

Heterogeneity: $Tau^2 = 0.1703$; $Chi^2 = 16.77$, df = 9 (P = 0.05); $I^2 = 46\%$

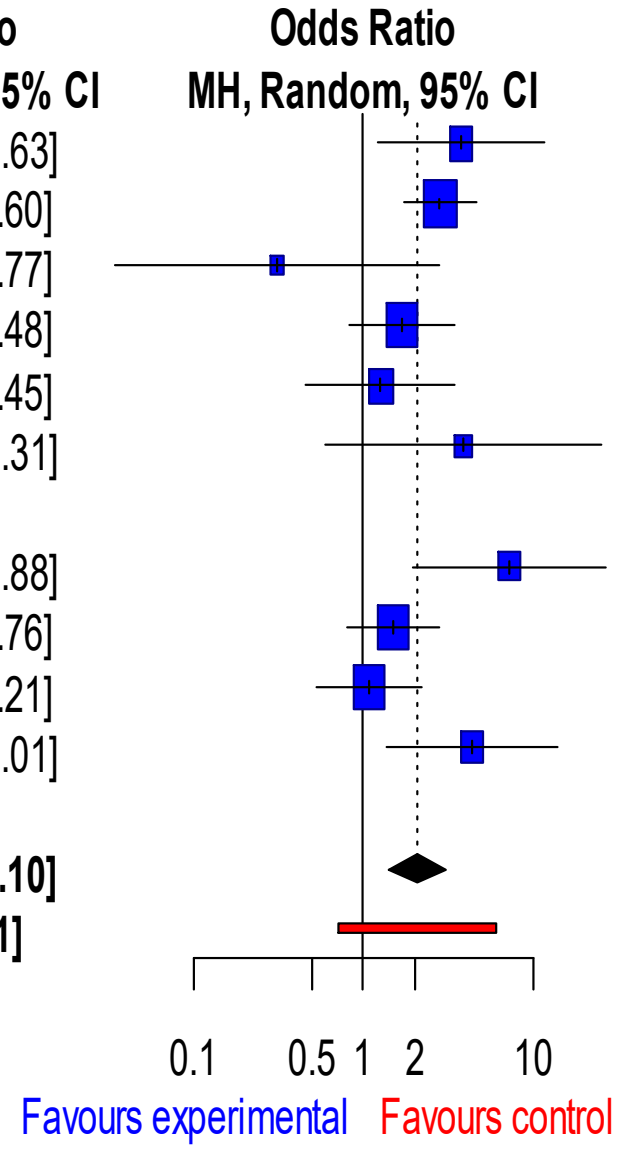

## L. CVD

| Study | Experimental Events | Experimental Total | Control Events | Control Total | Weight | Odds Ratio MH, Random, 95% CI |
|---|---|---|---|---|---|---|
| Wang et al. | 9 | 20 | 36 | 138 | 20.5% | 2.32 [0.89; 6.05] |
| Huang et al. | 3 | 2 | 13 | 41 | 0.0% | |
| Mo etal. | 14 | 15 | 55 | 155 | 6.0% | 25.45 [3.26; 198.77] |
| Chen et al. | 16 | 24 | 113 | 274 | 22.9% | 2.85 [1.18; 6.88] |
| Wang et al. | 21 | 53 | 65 | 339 | 34.2% | 2.77 [1.50; 5.11] |
| Chen TL et al. | 6 | 16 | 19 | 203 | 16.5% | 5.81 [1.90; 17.75] |
| **Total (95% CI)** | | **130** | | **1150** | **100.0%** | **3.46 [2.05; 5.87]** |
| **Prediction interval** | | | | | | **[0.88; 13.63]** |

Heterogeneity: $Tau^2 = 0.1131$; $Chi^2 = 5.88$, df = 4 (P = 0.21); $I^2 = 32\%$

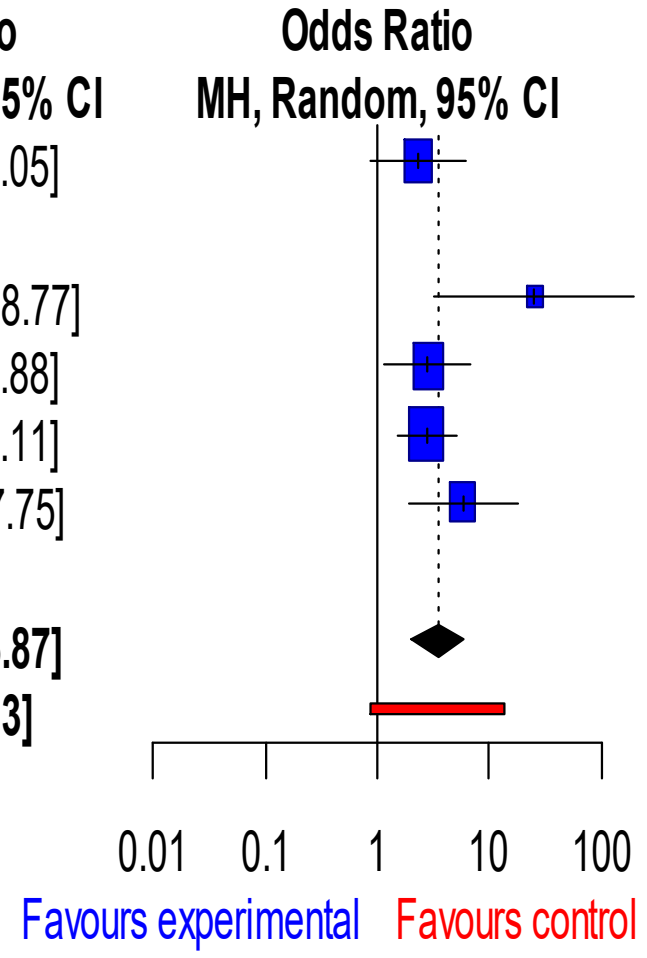

## M. Malignancy

| Study | Experimental Events | Experimental Total | Control Events | Control Total | Weight | Odds Ratio MH, Random, 95% CI |
|---|---|---|---|---|---|---|
| Wang et al. | 4 | 10 | 36 | 138 | 27.2% | 1.89 [0.50; 7.08] |
| Yang et al. | 1 | 1 | 52 | 52 | 0.0% | |
| Mo etal. | 5 | 7 | 55 | 155 | 17.0% | 4.55 [0.85; 24.21] |
| Chen et al. | 5 | 7 | 113 | 274 | 17.3% | 3.56 [0.68; 18.68] |
| Wang et al. | 3 | 15 | 65 | 339 | 28.4% | 1.05 [0.29; 3.84] |
| Chen TL et al. | 1 | 7 | 19 | 203 | 10.1% | 1.61 [0.18; 14.12] |
| **Total (95% CI)** | | **47** | | **1161** | **100.0%** | **2.04 [1.02; 4.07]** |
| **Prediction interval** | | | | | | **[0.67; 6.25]** |

Heterogeneity: $Tau^2 = 0$; $Chi^2 = 2.38$, df = 4 (P = 0.67); $I^2 = 0\%$

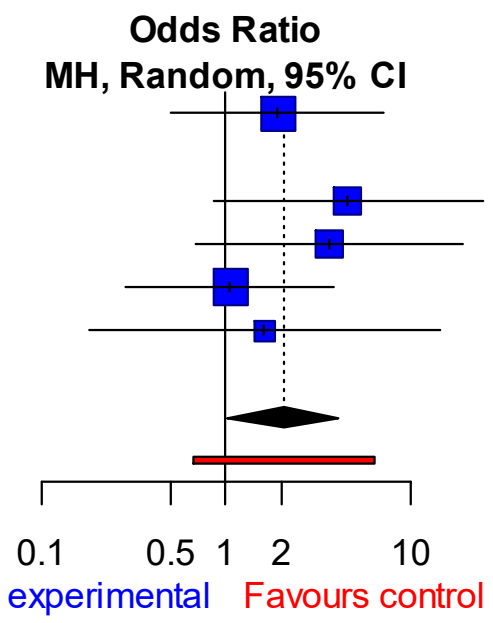

## N. COPD

| Study | Experimental Events | Experimental Total | Control Events | Control Total | Weight | Odds Ratio MH, Random, 95% CI |
|---|---|---|---|---|---|---|
| Wang et al. | 3 | 4 | 36 | 138 | 6.1% | 8.50 [0.86; 84.34] |
| Guan et al. | 6 | 12 | 173 | 1099 | 24.7% | 5.35 [1.71; 16.79] |
| Huang et al. | 1 | 1 | 13 | 41 | 3.0% | 6.33 [0.24; 165.88] |
| Zhang et al. | 2 | 2 | 58 | 140 | 3.5% | 7.05 [0.33; 149.60] |
| Liu j et al. | 3 | 5 | 17 | 61 | 9.2% | 3.88 [0.60; 25.31] |
| Mo etal. | 4 | 5 | 55 | 155 | 6.6% | 7.27 [0.79; 66.69] |
| Wang et al. | 11 | 21 | 65 | 339 | 40.0% | 4.64 [1.89; 11.38] |
| Chen TL et al. | 1 | 8 | 19 | 203 | 7.0% | 1.38 [0.16; 11.85] |
| **Total (95% CI)** | | **58** | | **2176** | **100.0%** | **4.76 [2.69; 8.39]** |
| **Prediction interval** | | | | | | **[2.34; 9.66]** |

Heterogeneity: $Tau^2 = 0$; $Chi^2 = 1.84$, df = 7 (P = 0.97); $I^2 = 0\%$

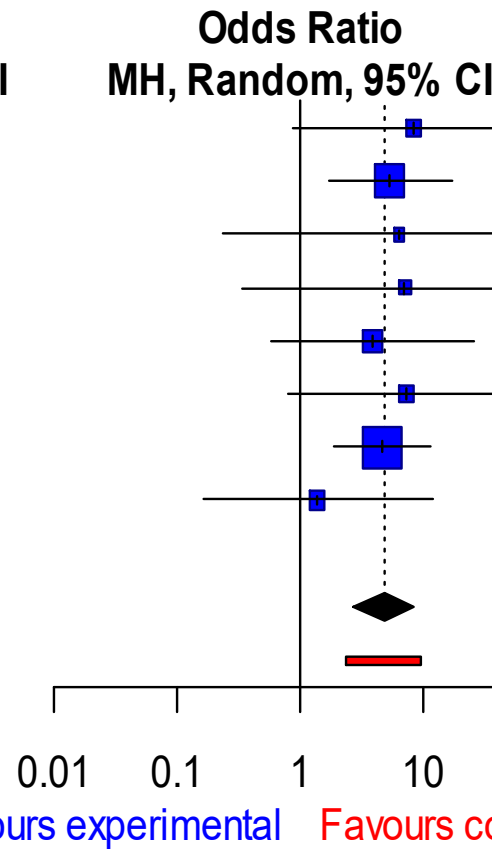

## O. CERD

| Study | Experimental Events | Experimental Total | Control Events | Control Total | Weight | Odds Ratio MH, Random, 95% CI |
|---|---|---|---|---|---|---|
| Wang et al. | 6 | 7 | 36 | 138 | 9.2% | 17.00 [1.98; 146.06] |
| Guan et al. | 4 | 15 | 173 | 1099 | 25.9% | 1.95 [0.61; 6.18] |
| Mo etal. | 7 | 7 | 55 | 155 | 5.3% | 27.16 [1.52; 484.56] |
| Chen et al. | 4 | 4 | 113 | 274 | 5.1% | 12.81 [0.68; 240.21] |
| Wang et al. | 10 | 21 | 65 | 339 | 36.6% | 3.83 [1.56; 9.41] |
| Chen TL et al. | 3 | 9 | 19 | 203 | 17.9% | 4.84 [1.12; 20.94] |
| **Total (95% CI)** | | **63** | | **2208** | **100.0%** | **4.54 [2.29; 8.99]** |
| **Prediction interval** | | | | | | **[1.16; 17.83]** |

Heterogeneity: $Tau^2 = 0.1215$; $Chi^2 = 5.97$, df = 5 (P = 0.31); $I^2 = 16\%$

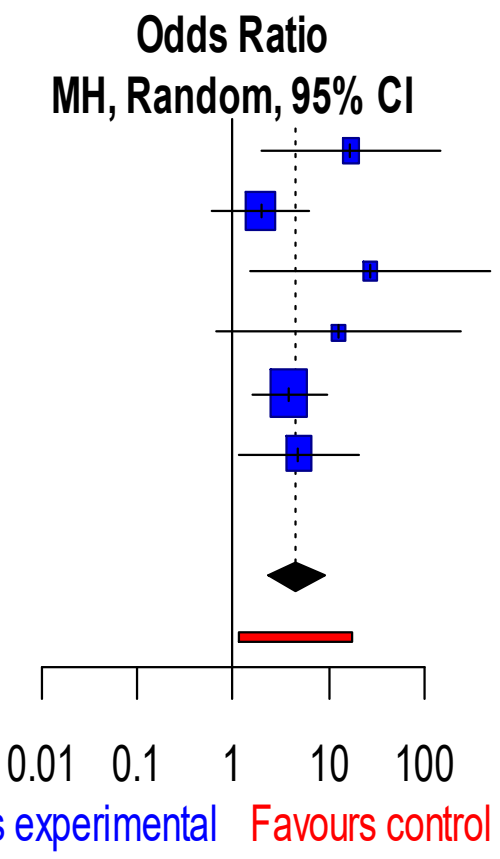

## P. CKD

| Study | Experimental Events | Experimental Total | Control Events | Control Total | Weight | Odds Ratio MH, Random, 95% CI |
|---|---|---|---|---|---|---|
| Wang et al. | 2 | 4 | 36 | 138 | 10.3% | 2.83 [0.38; 20.86] |
| Guan et al. | 3 | 8 | 173 | 1099 | 19.7% | 3.21 [0.76; 13.56] |
| Zhou et al. | 2 | 2 | 66 | 191 | 4.4% | 9.44 [0.45; 199.42] |
| Zhang et al. | 2 | 2 | 58 | 140 | 4.4% | 7.05 [0.33; 149.60] |
| Mo etal. | 4 | 6 | 55 | 155 | 13.7% | 3.64 [0.65; 20.49] |
| Chen et al. | 4 | 4 | 113 | 274 | 4.8% | 12.81 [0.68; 240.21] |
| Wang et al. | 4 | 13 | 65 | 339 | 28.0% | 1.87 [0.56; 6.27] |
| Chen TL et al. | 2 | 8 | 19 | 203 | 14.7% | 3.23 [0.61; 17.12] |
| **Total (95% CI)** | | **47** | | **2539** | **100.0%** | **3.22 [1.70; 6.10]** |
| **Prediction interval** | | | | | | **[1.45; 7.15]** |

Heterogeneity: $Tau^2 = 0$; $Chi^2 = 2.49$, df = 7 (P = 0.93); $I^2 = 0\%$

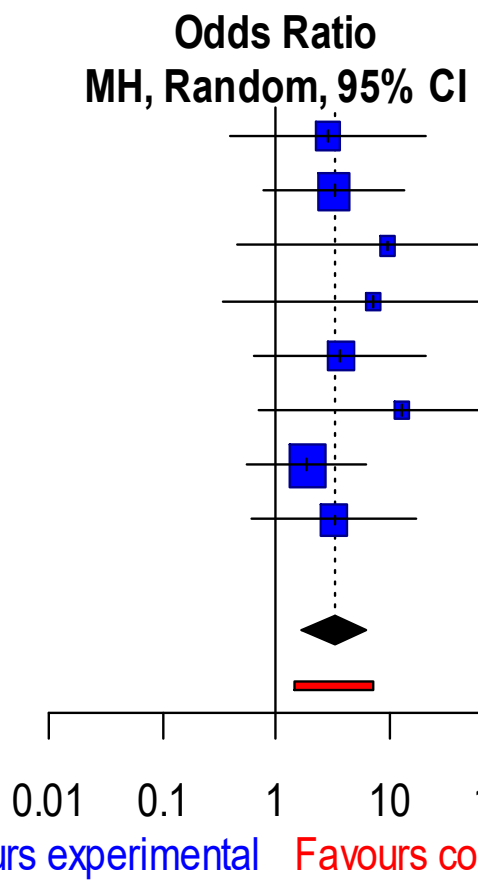

## Q. Smoking

| Study | Experimental Events | Experimental Total | Control Events | Control Total | Weight | Odds Ratio MH, Random, 95% CI |
|---|---|---|---|---|---|---|
| Guan et al. | 38 | 158 | 173 | 1099 | 68.5% | 1.69 [1.14; 2.53] |
| Zhou et al. | 5 | 11 | 66 | 191 | 7.3% | 1.58 [0.46; 5.37] |
| Zhang et al. | 6 | 9 | 58 | 140 | 5.4% | 2.83 [0.68; 11.77] |
| Liu j et al. | 2 | 4 | 17 | 61 | 2.6% | 2.59 [0.34; 19.87] |
| Mo etal. | 4 | 6 | 55 | 155 | 3.7% | 3.64 [0.65; 20.49] |
| Chen et al. | 9 | 19 | 113 | 274 | 12.6% | 1.28 [0.50; 3.26] |
| **Total (95% CI)** | | **207** | | **1920** | **100.0%** | **1.74 [1.25; 2.42]** |
| **Prediction interval** | | | | | | **[1.09; 2.78]** |

Heterogeneity: $Tau^2 = 0$; $Chi^2 = 1.74$, df = 5 (P = 0.88); $I^2 = 0\%$

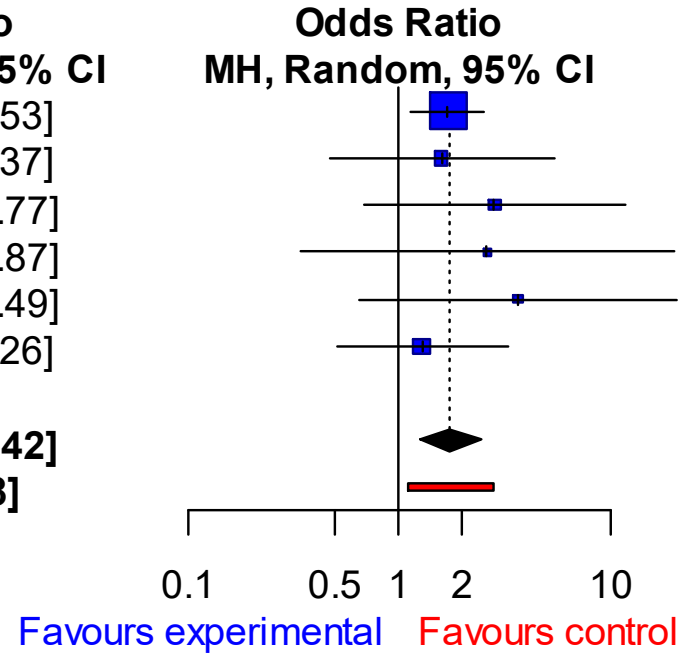

**Supplementary Figure 2: Meta-analysis of severity of comorbidities and symptoms in COVID-19 fatalities**

**A. Fever**

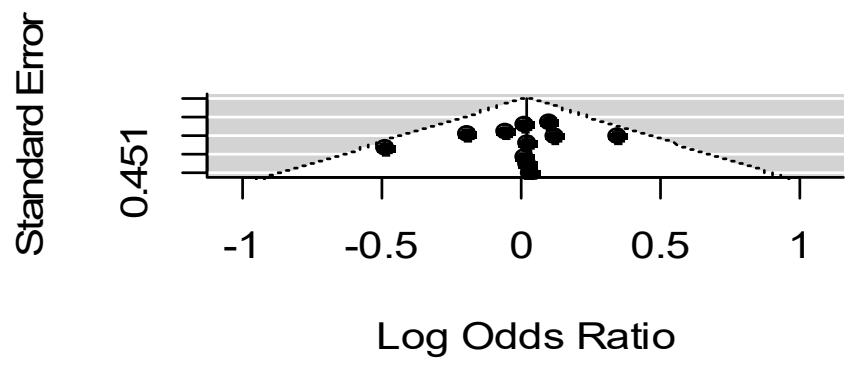

**Egger's test: p= 0.479**

**B. Cough**

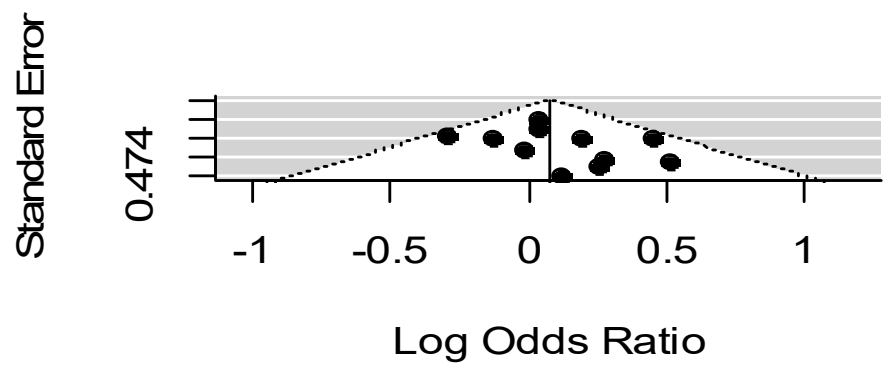

**Egger's test: p= 0.354**

**C. Fatigue**

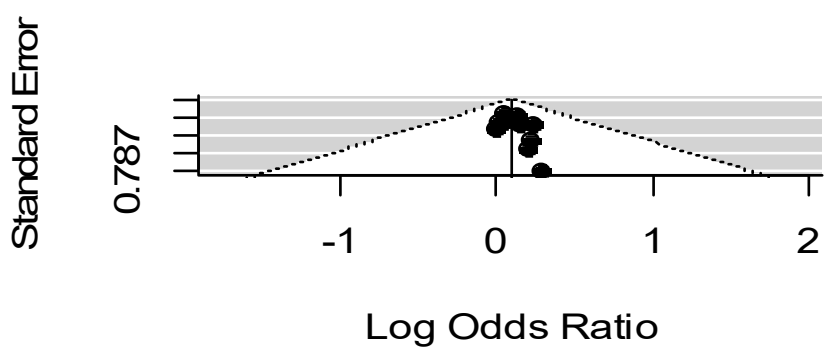

**Egger's test: p=0.183**

**D. Dyspnea**

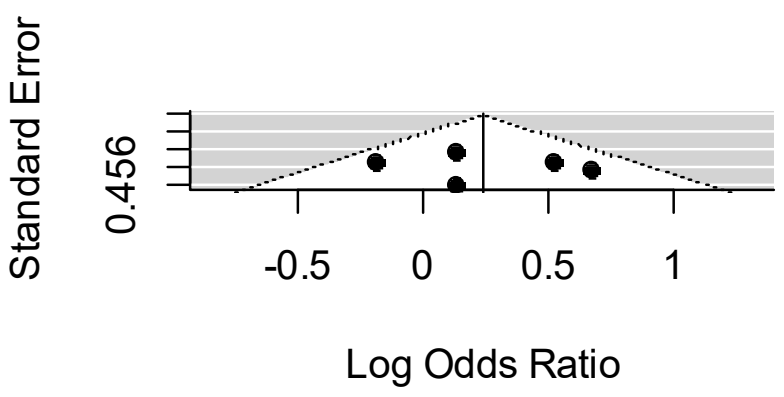

**Egger's test: p= 0.774**

**E. Myalgia**

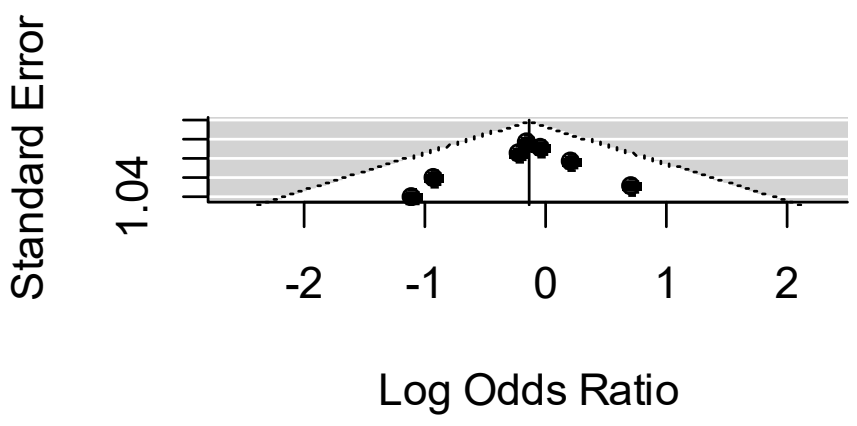

**Egger's test: p= 0.685**

**F. Anorexia**

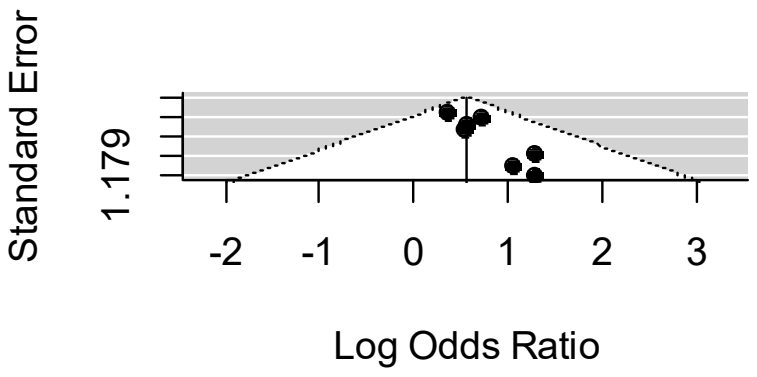

**Egger's test: p=0.018**

**G. Diarrhea**

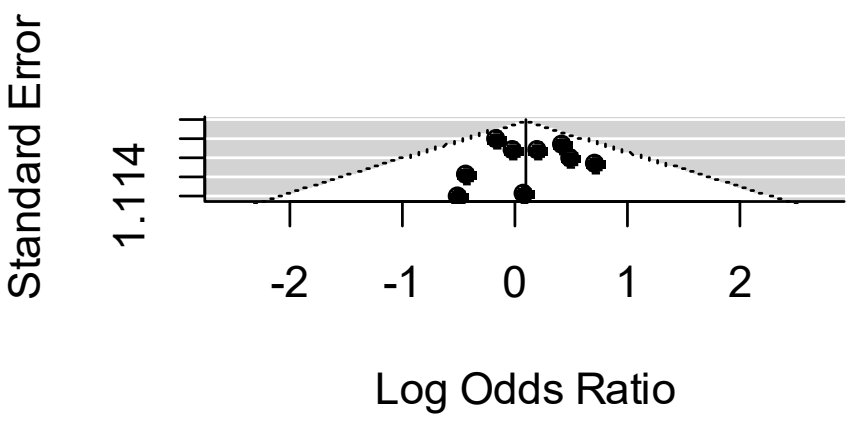

**Egger's test: p= 0.731**

**H. Nausea**

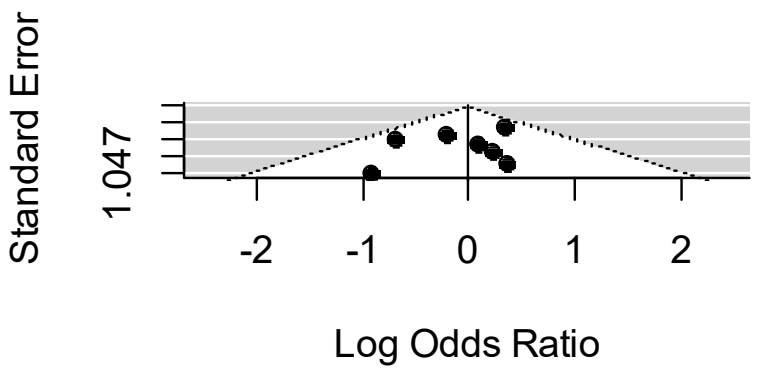

**Egger's test: p= 0.458**

**I. Headache**

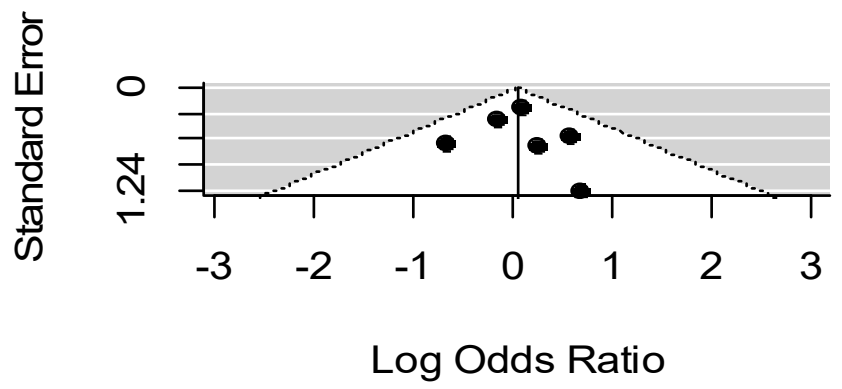

**Egger's test: p= 0.832**

**J. Hypertension**

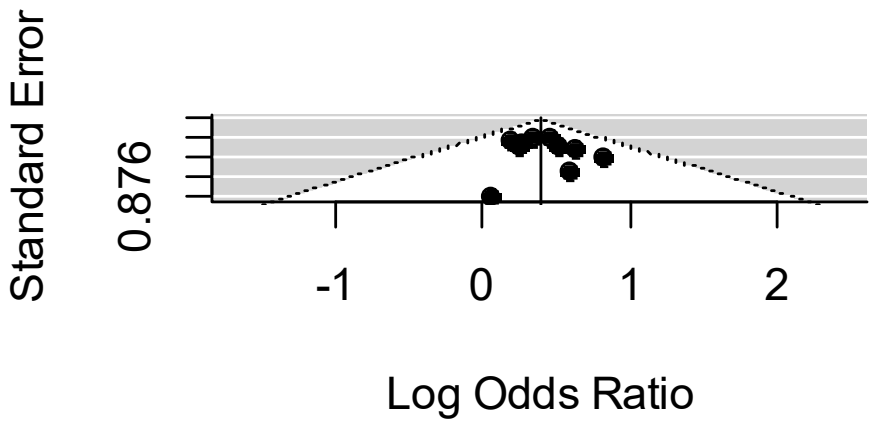

**Egger's test: p=0.551**

**K. Diabetes**

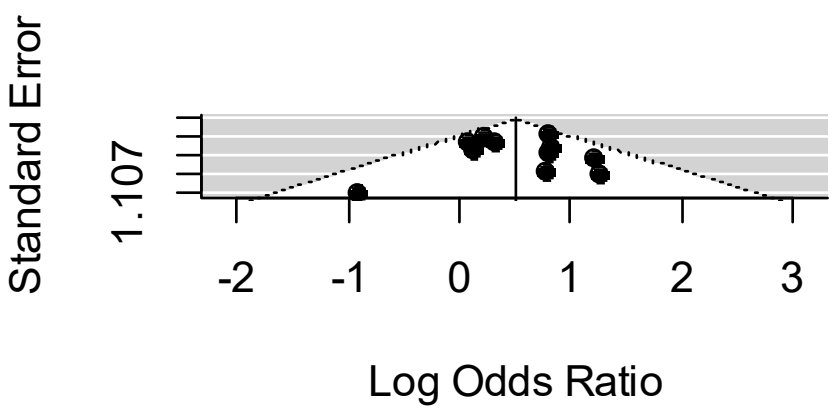

**Egger's test: p= 0.949**

**L. CVD**

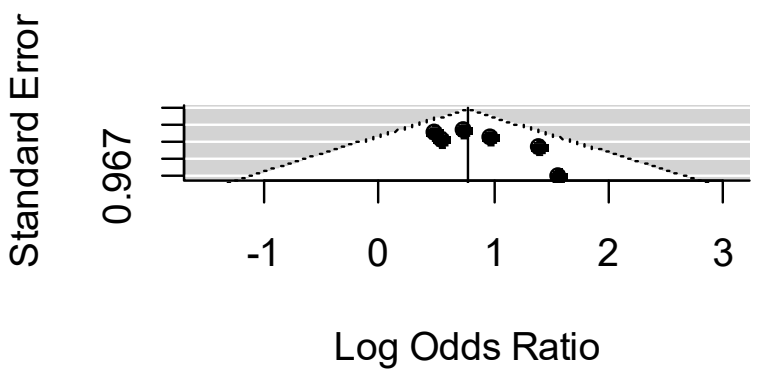

**Egger's test: p= 0.141**

**M. Malignancy**

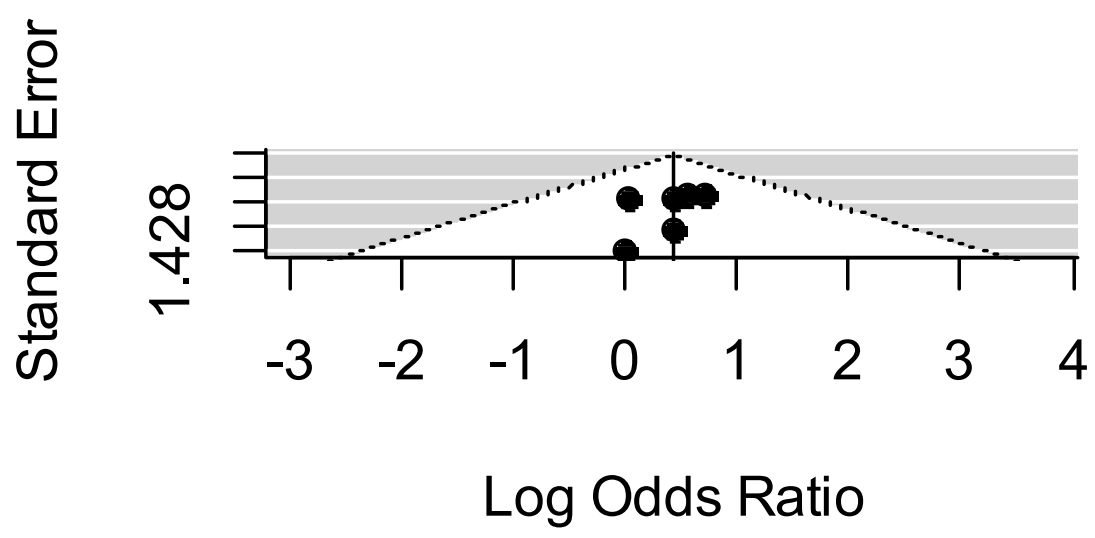

**Egger's test: p=0.466**

**N. CRED**

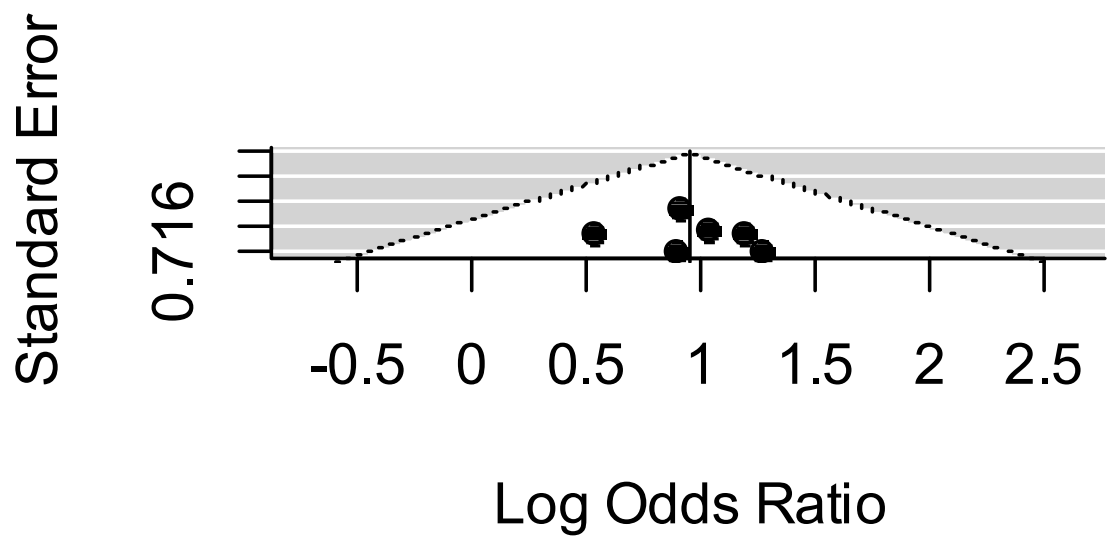

**Egger's test: p= 0.633**

**O. Smoking**

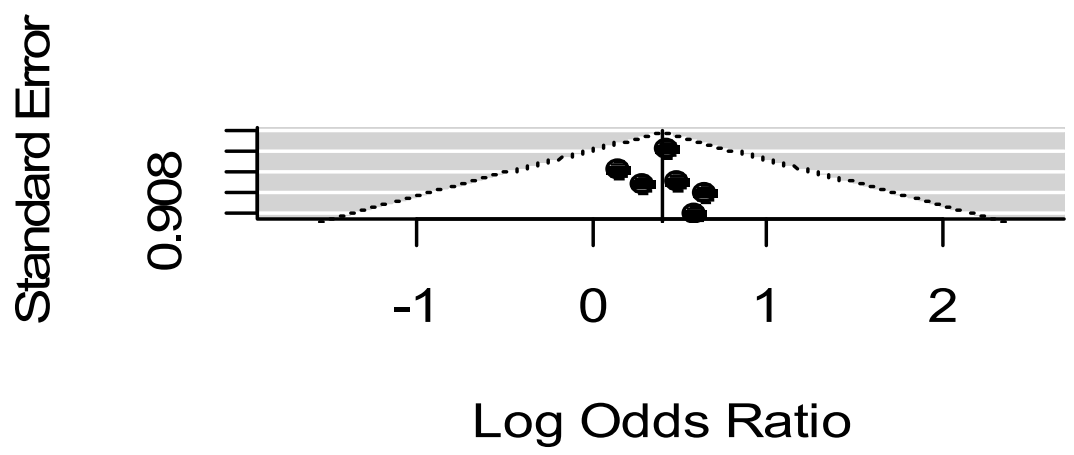

**Egger's test: p=0.916**

**P. CKD**

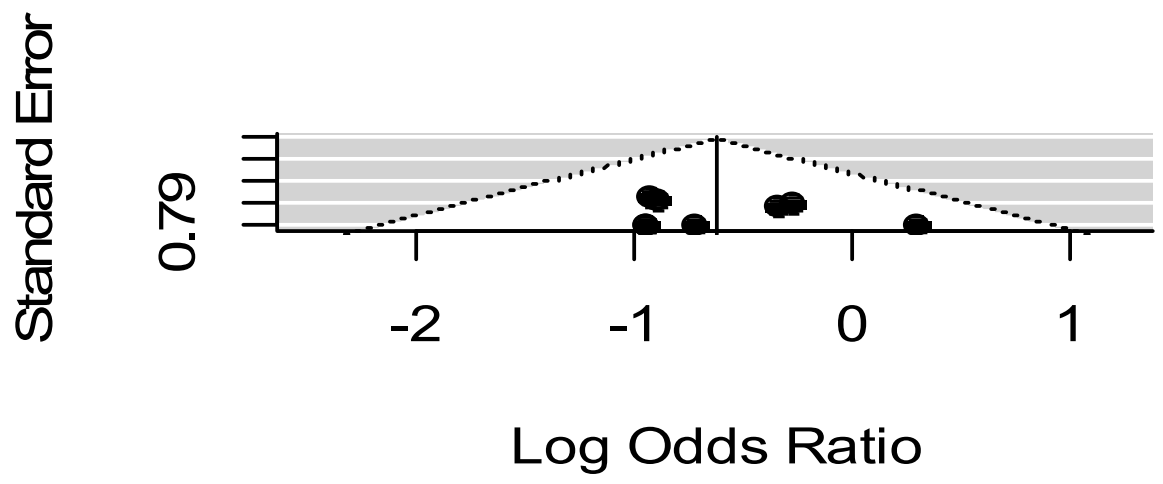

**Egger's test: p= 0.593**

**Q. COPD**

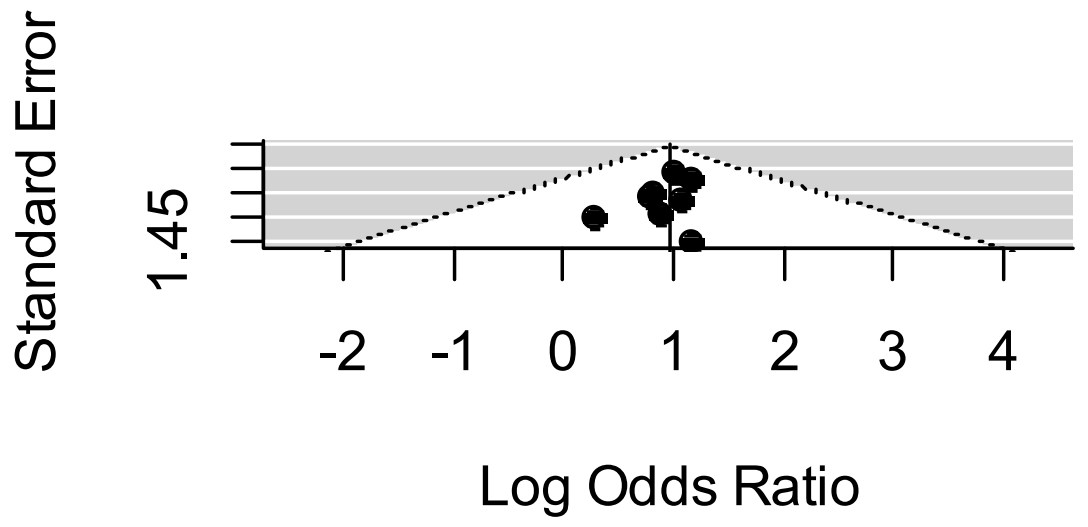

**Egger's test: p=0.235**

**Supplementary Figure 3: Assessment of publication bias using funnel plot and Egger's test**

**Supplementary Table 1: Accuracy and Evaluation matrices for symptoms data in ML analysis**

| Algorithms | Accuracy | Precision | Recall | f1 Score | AUC | Log loss |
|---|---|---|---|---|---|---|
| Random Forest | 0.87 | 0.89 | 0.93 | 0.91 | 0.83 | 4.52 |
| Decision Tree | 0.87 | 0.9 | 0.93 | 0.91 | 0.84 | 4.37 |
| XGB | 0.89 | 0.92 | 0.93 | 0.92 | 0.87 | 3.77 |
| GBM | 0.9 | 0.93 | 0.94 | 0.93 | 0.88 | 3.32 |
| SVM | 0.89 | 0.93 | 0.91 | 0.92 | 0.87 | 3.92 |
| Light GBM | 0.9 | 0.94 | 0.91 | 0.93 | 0.89 | 3.47 |

**Note: XGB= XGBoost; GBM= Gradient Boosting Machine; SVM= Support Vector Machine**

**Supplementary Table 2: Accuracy and Evaluation matrices for comorbidity data in ML analysis**

| Algorithms | Accuracy | Precision | Recall | F1 Score | AUC | Log loss |
|---|---|---|---|---|---|---|
| **Random Forest** | 0.87 | 0.89 | 0.93 | 0.91 | 0.82 | 4.52 |
| **Decision Tree** | 0.83 | 0.85 | 0.92 | 0.88 | 0.76 | 5.88 |
| **XGB** | 0.86 | 0.89 | 0.92 | 0.9 | 0.82 | 4.83 |
| **GBM** | 0.84 | 0.92 | 0.86 | 0.89 | 0.83 | 5.43 |
| **SVM** | 0.87 | 0.91 | 0.91 | 0.91 | 0.84 | 4.52 |
| **Light GBM** | 0.88 | 0.9 | 0.93 | 0.92 | 0.84 | 4.22 |

**Note: XGB= XGBoost; GBM= Gradient Boosting Machine; SVM= Support Vector Machine**

**Supplementary Table 3: Coefficient values for each symptom applying after ML methods**

| Algorithms | Headache | Fever | Cough | Fatigue | Nausea | Diarrhea | Myalgia | Dyspnea | Pneumonia | ARDS | Septic Shock |
|---|---|---|---|---|---|---|---|---|---|---|---|
| Random Forest | 0.3 | 1.89 | 1.74 | 0.61 | 0.18 | 0.16 | 0.71 | 2.94 | 76.62 | 4.77 | 0.69 |
| Decision Tree | 0.28 | 2.19 | 1.28 | 0.9 | 0.4 | 0.07 | 0.68 | 1.78 | 4.3 | 3.66 | 0.17 |
| XGB | 0 | 6.67 | 6.74 | 3.81 | 0 | 0 | 3.2 | 13.34 | 19.7 | 25.4 | 1.92 |
| GBM | 0.74 | 1.6 | 0.95 | 0.34 | 0.07 | 0.04 | 0.56 | 3.07 | 9.18 | 7.22 | 0.8 |
| SVM | 0 | 24.53 | 25.3 | 10.7 | 10.7 | 10.7 | 8.22 | 21.74 | 58.67 | 42.5 | 5.52 |
| Light GBM | 0 | 39.77 | 21.02 | 0 | 0 | 0 | 0 | 34.84 | 95.68 | 37.7 | 0.16 |

**Note: ARDS= Acute respiratory distress syndrome**

**Supplementary Table 4: Coefficient values for each comorbidity applying after ML methods**

| Algorithms | Gender | Age | Hypertension | CVD | CEVD | Chronic Lung Disease | Malignancy | Diabetes and Metabolic Disease | CLD | CKD | Neurodegenerative Disease | Infectious Disease | Surgical History | COPD | Asthma |
|---|---|---|---|---|---|---|---|---|---|---|---|---|---|---|---|
| **Random Forest** | 2.56 | 82.9 | 5.53 | 1.06 | 0.27 | 0.35 | 0.65 | 3.67 | 0.13 | 1.13 | 0.27 | 0.03 | 0.43 | 0.82 | 0.19 |
| **Decision Tree** | 4.05 | 85.51 | 2.57 | 0.49 | 0.28 | 0.45 | 0.88 | 3.77 | 0 | 0.81 | 0 | 0.007 | 0.55 | 0.52 | 0.1 |
| **XGB** | 9.37 | 20.77 | 14.84 | 5.98 | 0 | 2.9 | 7.37 | 14.21 | 0 | 7 | 0 | 0 | 0 | 5.77 | 11.8 |
| **GBM** | 2.19 | 77.78 | 6.48 | 1.37 | 0.27 | 0.23 | 0.73 | 7.72 | 0.09 | 0.87 | 0.46 | 0.05 | 0.64 | 1.02 | 0.09 |
| **SVM** | 4.07 | 169.3 | 34 | 10.18 | 23.89 | 11.66 | 0 | 33.58 | 22.7 | 29.2 | 14 | 15.78 | 8.34 | 25.4 | 24.15 |
| **Light GBM** | 54.6 | 1448.16 | 114.62 | 0.81 | 0 | 0 | 0 | 46.38 | 0 | 3.53 | 0 | 0 | 0 | 0 | 0 |

**Note: CVD= Cardiovascular disease; COPD=Chronic obstructive pulmonary disease; CEVD= Cerebrovascular disease; CKD=Chronic Kidney Disease; CLD= Chronic lung disease;**

**Supplementary Table 5: Assessing association between comorbidity and symptoms using Fisher's exact test of deceased patients**

| Comorbidity | Symptoms | P value | Comorbidity | Symptoms | P value |
|---|---|---|---|---|---|
| **Hypertension** | Headache | 0.232 | **CEVD** | Headache | 1.00 |
| | Fever | 0.423 | | Fever | 0.408 |
| | Cough | 0.823 | | Cough | 0.3183 |
| | Fatigue | 0.665 | | Fatigue | 1.00 |
| | Nausea or vomiting | 0.232 | | Nausea or vomiting | 1.00 |
| | Diarrhea | 1.00 | | Diarrhea | 1.00 |
| | Myalgia or arthralgia | 1.00 | | Myalgia or arthralgia | 0.03726 |
| | Dyspnea | 9.224e-06 | | Dyspnea | 1.00 |
| | Pneumonia | 2.2e-16 | | Pneumonia | 0.1905 |
| | ARDS | 3.22e-11 | | ARDS | 1.00 |
| | Septic Shock | 0.0006291 | | Septic Shock | 0.1868 |
| **Diabetes** | Headache | 0.179 | **CLD** | Headache | 1.00 |
| | Fever | 0.374 | | Fever | 1.00 |
| | Cough | 0.444 | | Cough | 0.008005 |
| | Fatigue | 0.613 | | Fatigue | 1.00 |
| | Nausea or vomiting | 1.00 | | Nausea or vomiting | 1.00 |
| | Diarrhea | 0.179 | | Diarrhea | 1.00 |
| | Myalgia or arthralgia | 0.447 | | Myalgia or arthralgia | 1.00 |
| | Dyspnea | 1.954e-05 | | Dyspnea | 0.02224 |
| | Pneumonia | 2.865e-11 | | Pneumonia | 1.00 |
| | ARDS | 1.638e-07 | | ARDS | 1.00 |
| | Septic Shock | 0.04661 | | Septic Shock | 1.00 |
| **CVD** | Headache | 1.00 | **Malignancy** | Headache | 1.00 |
| | Fever | 0.117 | | Fever | 1.00 |
| | Cough | 0.169 | | Cough | 1.00 |
| | Fatigue | 1.00 | | Fatigue | 0.08532 |
| | Nausea or vomiting | 1.00 | | Nausea or vomiting | 1.00 |
| | Diarrhea | 1.00 | | Diarrhea | 1.00 |
| | Myalgia or arthralgia | 1.00 | | Myalgia or arthralgia | 1.00 |
| | Dyspnea | 0.075 | | Dyspnea | 1.00 |
| | Pneumonia | 0.001414 | | Pneumonia | 0.1905 |
| | ARDS | 0.01666 | | ARDS | 0.5674 |
| | Septic Shock | 0.03891 | | Septic Shock | 0.1868 |

**Supplementary Table 5: Assessing association between comorbidity and symptoms using Fisher's exact test of deceased patients (continued…)**

| Comorbidity | Symptoms | P value | Comorbidity | Symptoms | P value |
|---|---|---|---|---|---|
| **Liver Disease** | Headache | 1.00 | **Surgical History** | Headache | 1.00 |
| | Fever | 1.00 | | Fever | 1.00 |
| | Cough | 0.008005 | | Cough | 1.00 |
| | Fatigue | 1.00 | | Fatigue | 0.02194 |
| | Nausea or vomiting | 1.00 | | Nausea or vomiting | 1.00 |
| | Diarrhea | 1.00 | | Diarrhea | 1.00 |
| | Myalgia or arthralgia | 1.00 | | Myalgia or arthralgia | 1.00 |
| | Dyspnea | 0.02224 | | Dyspnea | 1.00 |
| | Pneumonia | 1.00 | | Pneumonia | 1.00 |
| | ARDS | 1.00 | | ARDS | 1.00 |
| | Septic Shock | 1.00 | | Septic Shock | 1.00 |
| **COPD** | Headache | 0.01881 | **Asthma** | Headache | 1.00 |
| | Fever | 0.5457 | | Fever | 0.4813 |
| | Cough | 0.4382 | | Cough | 0.381 |
| | Fatigue | 0.1256 | | Fatigue | 1.00 |
| | Nausea or vomiting | 1.00 | | Nausea or vomiting | 1.00 |
| | Diarrhea | 1.00 | | Diarrhea | 1.00 |
| | Myalgia or arthralgia | 1.00 | | Myalgia or arthralgia | 1.00 |
| | Dyspnea | 0.2237 | | Dyspnea | 0.1645 |
| | Pneumonia | 0.01814 | | Pneumonia | 0.06186 |
| | ARDS | 0.08288 | | ARDS | 0.0002047 |
| | Septic Shock | 0.268 | | Septic Shock | 0.2281 |
| **Neurodegenerative Disease** | Headache | 1.00 | **CKD** | Headache | 0.05643 |
| | Fever | 0.01461 | | Fever | 0.7087 |
| | Cough | 0.008005 | | Cough | 1.00 |
| | Fatigue | 0.04347 | | Fatigue | 0.3367 |
| | Nausea or vomiting | 1.00 | | Nausea or vomiting | 1.00 |
| | Diarrhea | 1.00 | | Diarrhea | 1.00 |
| | Myalgia or arthralgia | 1.00 | | Myalgia or arthralgia | 1.00 |
| | Dyspnea | 0.02224 | | Dyspnea | 0.1642 |
| | Pneumonia | 1.00 | | Pneumonia | 1.67e-05 |
| | ARDS | 1.00 | | ARDS | 0.0003519 |
| | Septic Shock | 1.00 | | Septic Shock | 0.008497 |